\documentclass[aps,pra,11pt,final,longbibliography,onecolumn,showpacs,fleqn,eqsecnum,byrevtex,showkeys,floatfix,nofootinbib]{revtex4-1}

\usepackage{amsthm,amssymb,amsmath,amsxtra,stmaryrd,amscd}

\usepackage[a4paper]{geometry} 

\usepackage{graphicx} 

\usepackage{color} 
\usepackage[draft,color]{showkeys}
\definecolor{refkey}{rgb}{1,0,0}
\definecolor{labelkey}{rgb}{0,.7,.3}

\newcommand{\ie}{\textit{i.e.}}

\newcommand{\Rset}{{\mathbb{R}}}

\newcommand{\Mmc}{{\mathcal{M}}}
\newcommand{\Em}{{\mathcal{E}}}
\newcommand{\Us}{\mathcal{U}}
\newcommand{\Ccl}{{\mathcal{C}}}

\newcommand{\Frap}{{\mathfrak{F}}}

\newcommand{\Ptcl}{{\widetilde{\mathcal{P}}}}
\newcommand{\Emt}{{\widetilde{\mathcal{E}}}}
\newcommand{\Cclt}{\widetilde{\mathcal{C}}}

\newcommand{\Evt}{{\widetilde{E}}}
\newcommand{\Pt}{{\widetilde{P}}}
\newcommand{\Wt}{{\widetilde{W}}}
\newcommand{\Wh}{{\widehat{W}}}

\newcommand{\Emh}{{\widehat{\mathcal{E}}}}
\newcommand{\Phcl}{{\widehat{\mathcal{P}}}}
\newcommand{\Cclh}{\widehat{\mathcal{C}}}

\newcommand{\taut}{{\tilde{\tau}}}
\newcommand{\kt}{{\tilde{k}}}

\newcommand{\Cclb}{\overline{\mathcal{C}}}
\newcommand{\Emb}{{\overline{\mathcal{E}}}}

\newcommand{\Evb}{{\overline{E}}}

\newcommand{\tauh}{{\hat{\tau}}}
\newcommand{\kh}{{\hat{k}}}

\newcommand{\Evh}{{\widehat{E}}}
\newcommand{\Ph}{{\widehat{P}}}

\newcommand{\Cscc}{{\mathcal{C}}}
\newcommand{\EmS}{{\mathcal{S}}}
\newcommand{\Pcl}{{\mathcal{P}}}

\newcommand{\Cclr}{\mathring{\mathcal{C}}}
\newcommand{\Emr}{{\mathring{\mathcal{E}}}}
\newcommand{\Evr}{{\mathring{E}}}
\newcommand{\taur}{{\mathring{\tau}}}

\newcommand{\taub}{{\bar{\tau}}}

\begin{document}

\title{Relativistic Stereometric Coordinates from Relativistic Localizing Systems and the Projective Geometry of the Spacetime Manifold}

\author{Jacques \surname{L.} Rubin}

\affiliation{
(email: \texttt{jacques.rubin@inln.cnrs.fr}) \\
Universit{\'e} de Nice--Sophia Antipolis, UFR Sciences\\
Institut du Non-Lin{\'e}aire de Nice, UMR7335\\ 
1361 route des Lucioles,
F-06560 Valbonne,
France}

\date{\today}

\begin{abstract}
Relativistic stereometric coordinates supplied by relativistic auto-locating positioning systems made up of four satellites supplemented by a fifth one are defined in addition to the well-known emission and reception coordinates. Such a constellation of five satellites defines a so-called relativistic localizing system. The determination of such systems is motivated by the need to not only locate (within a grid) users utilizing receivers but, more generally, to localize any spacetime event.  The angles measured on the celestial spheres of the five satellites enter into the definition. Therefore, there are, up to scalings, intrinsic physical coordinates related to the underlying conformal structure of spacetime. Moreover, they indicate that spacetime must be endowed everywhere with a local projective geometry characteristic of a so-called generalized Cartan space locally modeled on four-dimensional, real projective space. The particular process of localization providing the relativistic stereometric coordinates is based, in a way, on an enhanced notion of parallax in space and time generalizing the usual parallax restricted to space only.
\end{abstract}


\maketitle

\section{A protocol implemented by users to localize events \label{intro}}

Almost simultaneously, Bahder \cite{bahder2001navigation}, Blagojevi{\'c} \textit{et al.} \cite{Blago:2002}, Coll \cite{coll2013relativistic}
 and Rovelli \cite{Rovelli:2002gps} laid, from different approaches, the foundations of the relativistic positioning systems (RPS) and, in particular, following Coll's terminology,\footnote{ %
Throughout the present paper, we use terms such as \textit{primary}, \textit{local}, \textit{intrinsic}, \textit{location system}, \textit{reference system}, \textit{positioning system}, \textit{auto-locating system}, \textit{autonomous system} or \textit{data}, \textit{laws of physics}, \textit{emission} and \textit{reception coordinates}, etc., as defined in \cite{coll2013relativistic}
} %
``primary'' RPSs, \ie, RPSs which satisfy the three following criteria: they are 1) ``generic,'' \ie, the system of coordinates they provide must exist independently of the spacetime geometry for each given class of spacetime, 2) they are ``free,'' \ie, their structures do not need the knowledge of the gravitational field, and 3) they are ``immediate,'' \ie, the users know their positions without delay at the instant they receive the four ``time stamps $\tau_\alpha$'' sent by the four emitting satellites of the RPS satellite constellation. 
\par
Among this set of primary RPSs, there exists the sub-class of the so-called ``auto-locating RPSs,'' \ie, those RPSs in which each satellite broadcasts its own time stamp but also the time stamps it receives from its neighboring satellites. The SYPOR system (``\textit{SYst{\`e}me de POsitionnement Relativiste}''), developed by Coll and Tarantola \cite{Coll:2003fj}, belongs to this category, but we ask, more generally, for an enhanced RPS and a supplementary protocol to allow any located user to localize
any event in the spacetime region covered by this particular enhanced RPS. 
\par
We make the following strict distinction between \textit{location} and \textit{localization}. To locate an event, a protocol (of location) is needed to build a coordinate grid, and then, to position this event in this grid once the coordinates of this event are known. To localize an event, a protocol (of localization) is needed that effectively obtains the coordinates of the event to be then, only, located in a given coordinate grid. Auto-locating positioning systems only allow building the coordinate grids from the users' knowledge of the satellites' worldlines, and then, to position the users in these grids, but they do not supply the coordinates of events. Upstream, non auto-locating systems only allow knowing the users' coordinates but without location and, more generally, without localization of events in the users' surroundings.
\par
Furthermore, downstream, the sub-class of the so-called ``autonomous systems,'' contained in the sub-class of auto-locating systems, includes those auto-locating systems allowing, from ``autonomous data,'' the users to draw (from Coll's definition \cite{coll2013relativistic}) the satellites' worldlines in the spacetime where these users are living. Beside, we consider rather another sub-class contained in the sub-class of auto-locating systems, namely, the sub-class of ``relativistic localizing systems'' of which the satellites broadcast also, in addition to their time stamps, data to localize events. In the present paper, we define such a relativistic localizing system made up of four satellites constituting an auto-locating system supplemented by an ancillary fifth satellite providing data (actually, supplementary time stamps) to localize events. These five satellites can define five different auto-locating systems connected by ten changes of coordinate grids but only one of the five is required to operate.
\par\medskip
Besides, the goal for seeking such an enhanced RPS, \textit{viz.}, a relativistic localizing system, provided with a tracking, localizing protocol is also to find a process to break the underlying arbitrariness in scaling that is due, in a way, to the arbitrary choice of time parameterizations of the satellites' worldlines. Indeed, the satellites of a given RPS satellite constellation can broadcast time stamps defined not only by their own proper times given by on-board clocks, but, more generally, by any ``numbered events generator'' (such as proper time clocks) not necessarily synchronized with their proper times. Thus, any time parameterization can be defined, in particular, affinely from any other given time parameterization. In other words, the links between, on the one hand, the conformal structure of spacetime with, behind it, the time parameterization scalings and, on the other hand, the time parameterizations of the satellites' worldlines must be questioned, technologically unveiled, and then fixed by a particular enhanced RPS. By ``fixed,'' we mean that the enhanced RPS should be ``sensitive'' to the conformal structure of spacetime and then, in particular, sensitive to any scaling change of the Lorentzian metric defined on spacetime. But, it should also provide a univocal linkage with the conformal structure and, in addition, this linkage must be unaffected by the changes in the time parameterization along the satellites' worldlines. 
\par\medskip
Furthermore, the conformal structure of spacetime can be deduced from the causal axiomatics as shown, historically, for instance, by Ehlers \textit{et al.} \cite{EhlPiSch:72}, Hawking \textit{et al.} \cite{hawk76}, Kronheimer \textit{et al.} \cite{kronpen67}, Malament \cite{malam77}, or Woodhouse \cite{wood73}. As a consequence, the chronological order, \ie, the history in spacetime, is not affected by scalings of the Lorentzian metric. Hence, the changes of coordinates in spacetime which are compatible with the chronological order transform the Lorentzian metric up to scalings, \ie, up to functional conformal factors. 
Then, the Lorentzian metric is said to be ``conformally equivariant.'' 
As a consequence of this conformal structure, only the generators of the null cones and not their constitutive sets of points (events) are then the intrinsic, hybrid, and causal objets intertwining physics and geometry that should be used in the geometrical statements of the laws of physics. And then the events should be only considered as the intersection points of congruences of such generators.
\par
Hence, intrinsic (physical) observables and ``genuine, causal processes'' such as the location protocols must be unaffected by metric scalings, \ie, metric scalings are not intrinsic. Also, coordinate systems such as emission or reception coordinates which can be subjected to scalings due to changes of time parameterization of, for instance, the satellites' worldlines, are then also not intrinsic.
Therefore, we must, somehow, discriminate in any given coordinate system its intrinsic part from its ``scaling sensitive,'' non-intrinsic part. Actually, an auto-locating system cannot provide such a discrimination, as will be shown in what follows. A fifth satellite must be attached to this positioning system. Using a metaphor, this fifth satellite is a sort of cursor indicating the scale of the positioning system from which an intrinsic part alone can be excerpted. Moreover, this intrinsic part cannot provide by itself a complete, functional coordinate system.
\par
Angles on a celestial sphere are such intrinsic observables compatible with the conformal structure of spacetime. But, their evaluations
from a causal (intrinsic) process of measurement need a particular protocol if an auto-locating system only is involved. Such a protocol is presented in the next sections using emission coordinates with a fifth satellite. In return,  we obtain, from the emission coordinates provided by this particular five-satellite constellation, a local relativistic localizing system defined with new coordinates, namely, the ``relativistic stereometric coordinates.'' As a result, we gain much more than a ``mere'' auto-locating system with a fifth satellite since not only location is then available but localization, in addition, becomes available. Also, a ``stereometric grid'' is obtained and linked to the emission grid provided by the auto-locating sub-system. Furthermore, it appears that spacetime must be embedded in a five-dimensional, intermediate manifold in which spacetime must be considered locally as a four-dimensional, real projective space, \ie, spacetime is then a \textit{generalized Cartan space} ``modeled'' on a projective space. Thus, we obtain a local, projective description of the spacetime geometry. Nevertheless,  we have, in return, access to the ``genuine'' Riemannian four-dimensional spacetime structure without the need for any autonomous sub-system unless considering that the five-satellite constellation constitutes a sort of ``enhanced autonomous system.'' This kind of protocol can be called a \textit{relativistic stereometric protocol} \cite{coll2013relativistic}.
\par
In the next sections, we present such a complete protocol. It has two major flaws which we nevertheless think are unavoidable: its implementation is complicated and may be immediate only in some very particular situations or regions covered by the RPS depending on the localized events. In full generality, obviously, it cannot be immediate, because the satellites of any constellation must ``wait'' for the signals coming from the source event which will be later localized. Nevertheless, it really breaks the scaling arbitrariness and provides access to the spacetime $\Mmc$ as expected. Moreover, it may possibly give a completely new interpretation of a particular sort of the so-called ``Weyl's length connection'' which may circumvent, by construction, the fundamental criticisms made by Einstein. 
\par\medskip
The results presented in the next sections are given when increasing successively the dimension of spacetime. Thus, in Section~\ref{sec2D}, the relativistic localizing protocol is applied in a two-dimensional spacetime. In this particular case only, the relativistic localizing system essentially reduces to the relativistic positioning system itself. In Section~\ref{sec3D}, all of the basic grounds and principles of the localizing process are presented in a three-dimensional spacetime. Then, they are naturally generalized in Section~\ref{sec4D}  to a four-dimensional spacetime before ending with the conclusion in Section~\ref{conclusion}.

\section{The protocol of localization in a $(1+1)$-dimensional spacetime $\Mmc$ \label{sec2D}}

\begin{figure}[ht] 
\centering 
\includegraphics[width=.55\columnwidth]{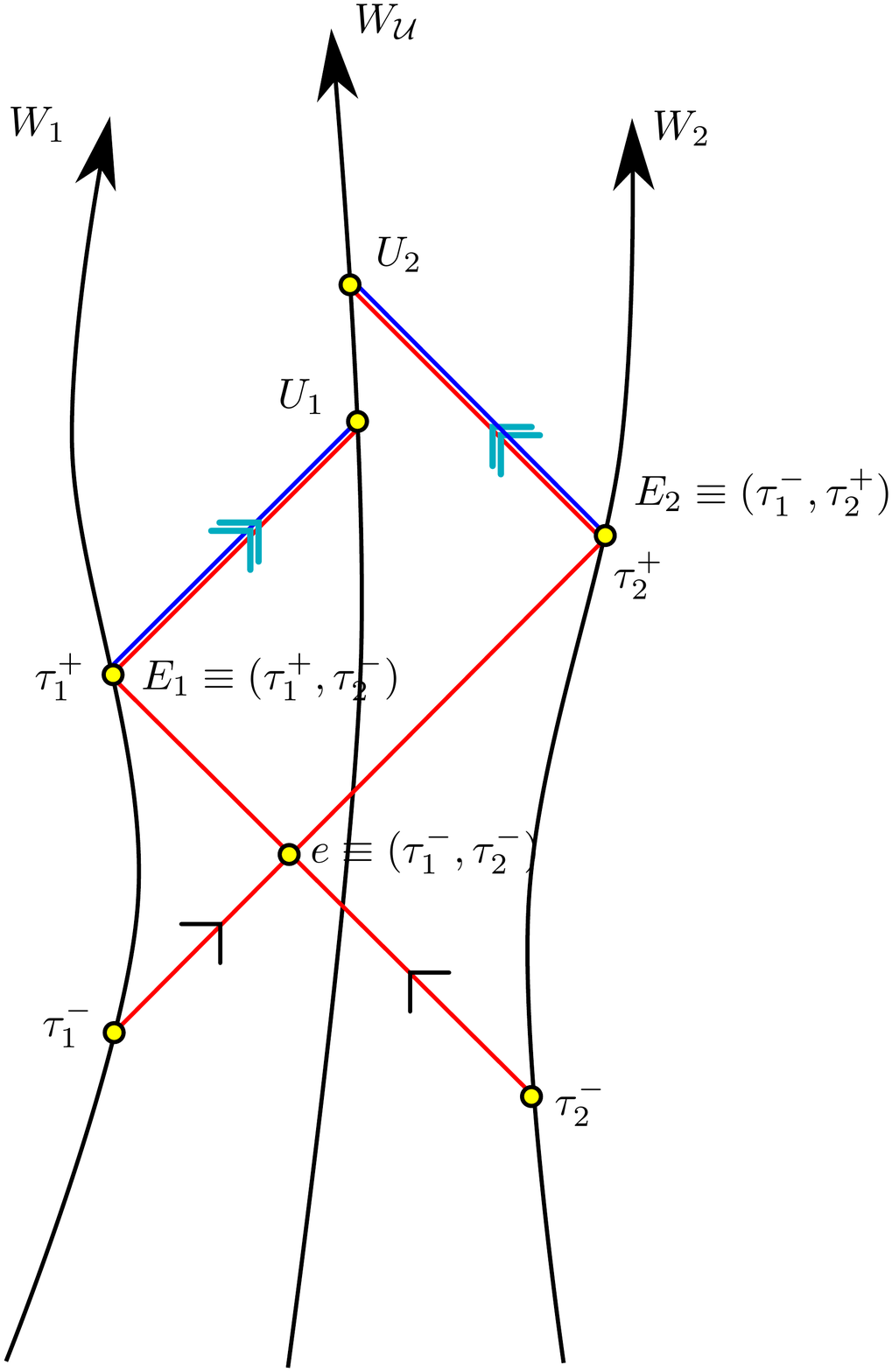}
\caption{\footnotesize \quad The system of echoes in a two-dimensional spacetime.}
\label{figechoe2D}
\end{figure}

\begin{figure}[ht] 
\centering 
\includegraphics[width=.7\columnwidth]{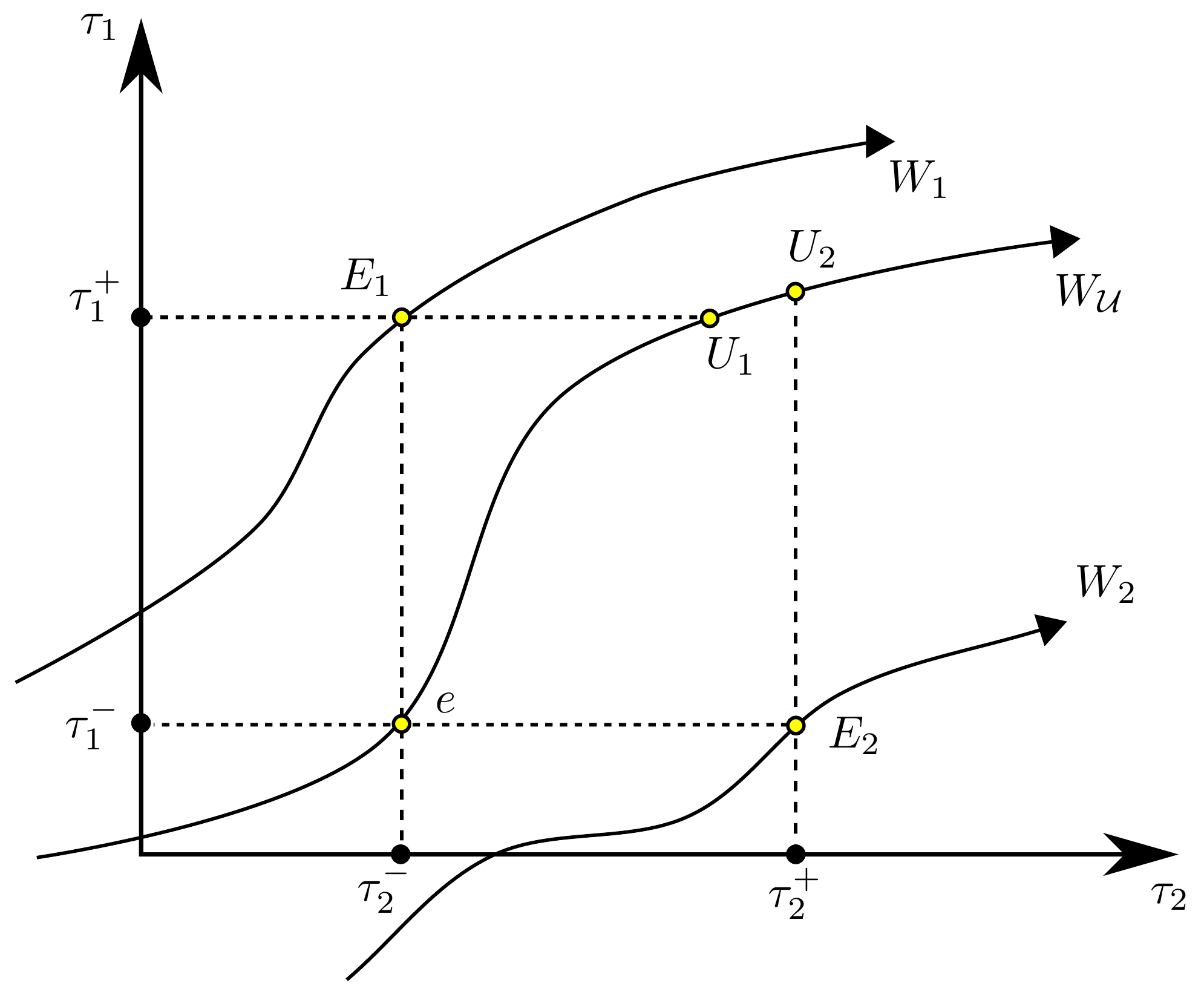}
\caption{\footnotesize \quad The two-dimensional emission/stereometric grid.}
\label{figgrid2D}
\end{figure}

In this situation, the protocol is rather simple. We recall, first, the principles for  relativistic positioning with a two-dimensional auto-locating system. We consider two emitters, namely, $\Em_1$ and $\Em_2$ and a user $\Us$ with their respective (time-like) worldlines  $W_1$, $W_2$ and $W_\Us$. The two emitters broadcast\textit{ emission coordinates} which are two time stamps $\tau_1$ and $\tau_2$ generated by on-board clocks, and then the two-dimensional \textit{emission grid} can be constructed from this RPS. From a system of echoes (Fig.~\ref{figechoe2D}), the user at the events $U_1\in W_\Us$ and $U_2\in W_\Us$ receives four numbers: $(\tau_1^+,\tau_2^-)$ from $E_1$ and 
$(\tau_1^-,\tau_2^+)$ from $E_2$ (see Fig.~\ref{figgrid2D}).

In addition, from this RPS, the user can also know in this emission grid the two events $E_1$ and $E_2$ at which the two emitters sent these four time stamps \textit{viz}, $E_1\equiv(\tau_1^+,\tau_2^-)$ and $E_2\equiv(\tau_1^-,\tau_2^+)$.
\par
Then, let $e$ be an event in the domain covered by the RPS (mainly between the two worldlines $W_1$ and $W_2$). This event can be at the intersection point of the two light rays received by $\Em_1$ and $\Em_2$ at the events $E_1$ and $E_2$ (see Fig.~\ref{figgrid2D}). Hence, the position of $e$ in the emission grid is easily deduced by $\Us$ if 1) $\Us$ records $(\tau_1^+,\tau_2^-)$ and $(\tau_1^-,\tau_2^+)$ along $W_\Us$, and 2) a physical identifier for $e$ is added at $E_1$ and $E_2$ to each pair of time stamps to be matched by $\Us$. 

Thus, in the particular case of two dimensions, the emission grid is identified with the stereometric grid and, as a consequence, the stereometric coordinates are also identified  with the emission coordinates. 

\section{The protocol of localization in a $(2+1)$-dimensional spacetime $\Mmc$ modeled on $\Rset P^3$  \label{sec3D}}

\begin{figure}[ht] 
\centering 
\includegraphics[width=\columnwidth]{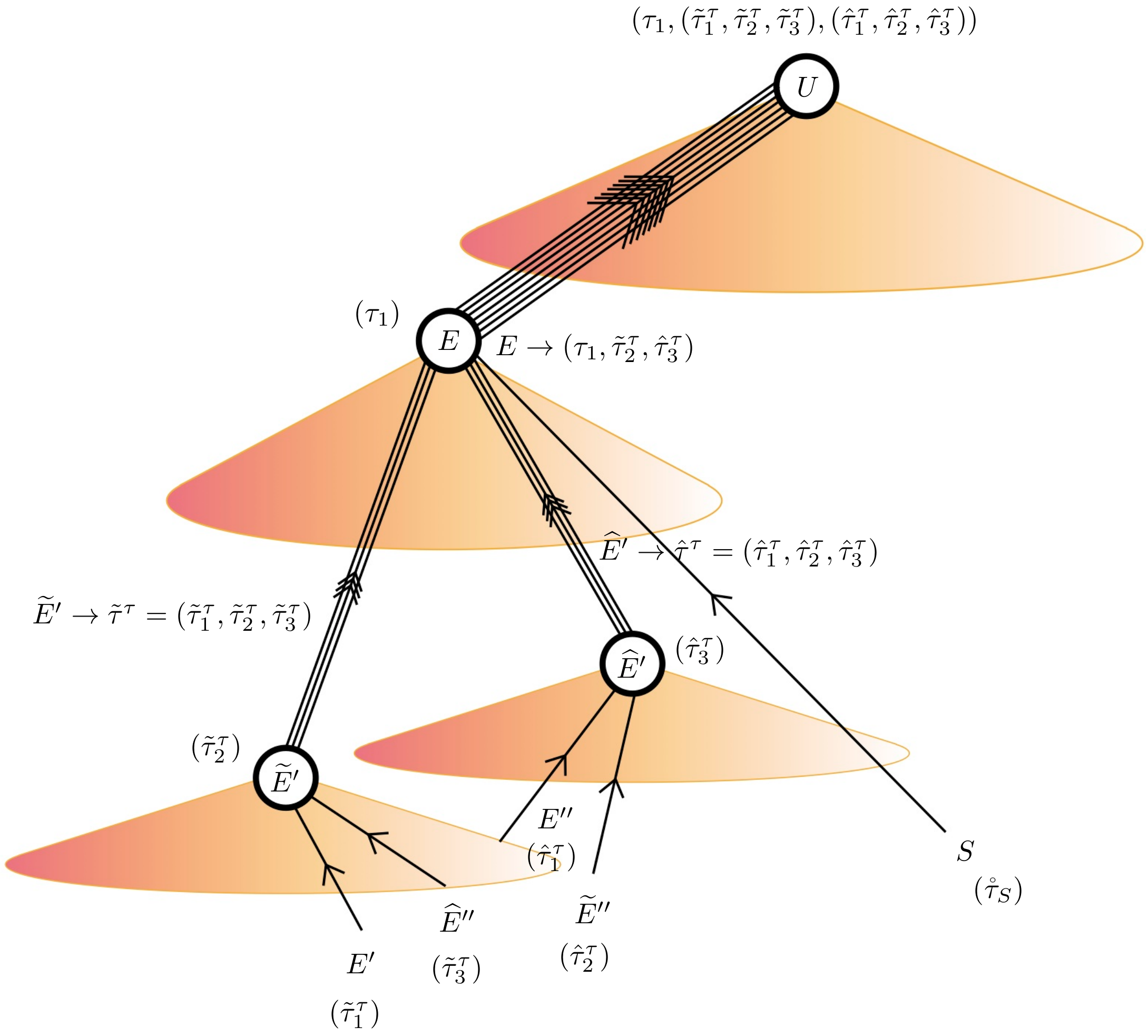}
\caption{\footnotesize\quad The system of echoes with four past null cones.}
\label{figechoes3D}
\end{figure}

In this case, the complexity of the protocol of localization increases ``dramatically.'' Again, we consider three emitters $\Em$, $\Emt$ and $\Emh$ transmiting three sets of time stamps denoted, respectively, by $\tau$, $\taut$ and $\tauh$. Then, the emission grid is the Euclidean space $\Rset^3$ with the system of Cartesian emission coordinates 
$(\tau,\taut,\tauh)$. Then, we consider, first, the system of echoes from $\Em$ to the user $\Us$. This system can be outlined as indicated in Fig.~\ref{figechoes3D}. 

In this figure, the four past null cones of the four events $E$, $\Evt'$, $\Evh'$ and $U$ are represented and the time axis is oriented vertically from the bottom to the top of the figure. Also, we denote by ${}^{\Us}W$, $W$, $\Wt$ and $\Wh$ the worldlines of, respectively, the user $\Us$ and  the emitters $\Em$, $\Emt$ and  $\Emh$. 
\par
Then, the user receives at the reception event $U\in {}^{\Us}W$ seven time stamps sent by $\Em$ and emitted at the event of emission $E\in W$: $(\tau_1,(\taut^\tau_1,\taut^\tau_2,\taut^\tau_3),(\tauh^\tau_1,\tauh^\tau_2,\tauh^\tau_3))$.
In addition, the emitter $\Em$ receives at $E$ six time stamps from the other two emitters $\Emt$ and $\Emh$, \textit{viz}, $p_{\Evt'}\equiv(\taut^\tau_1,\taut^\tau_2,\taut^\tau_3)$ emitted at $\Evt'\in\Wt$ from $\Emt$, and $p_{\Evh'}\equiv(\tauh^\tau_1,\tauh^\tau_2,\tauh^\tau_3)$ emitted at $\Evh'\in\Wh$ from $\Emh$. Actually, $p_{\Evt'}$ and $p_{\Evh'}$ are the 3-positions of, respectively, $\Evt'$ and $\Evh'$ in the three-dimensional emission grid. 
Moreover, $\Em$ sends at $E$ the time stamp $\tau_1$ received at $U$ by the user $\Us$. 
\par
In addition, two of the three time stamps received at $\Evt'$ are sent by $\Em$ at $E'$: $\taut^\tau_1$, and by $\Emh$ at $\Evh''$: $\taut^\tau_3$; and we have a similar situation for $\Evh'$ (see Fig.~\ref{figechoes3D}). 
\par
The user can then deduce the 3-position $p_E$ of the event $E$ in the emission grid: $p_E\equiv(\tau_1,\tau_2,\tau_3)\equiv(\tau_1,\taut^\tau_2,\tauh^\tau_3)$, and the two 3-positions $p_{\Evt'}$
and $p_{\Evh'}$ of the two events $\Evt'$ and $\Evh'$ respectively. In addition, $\taut^\tau_2$ is emitted by $\Emt$ at $\Evt'$, and $\tauh^\tau_3$ is emitted by $\Emh$ at $\Evh'$.
Also, these two 3-positions are obtained from four time stamps emitted from four events, namely, $E'$ and $\Evh''$ for $\Evt'$, and $E''$ and $\Evt''$ for $\Evh'$ (see Fig.~\ref{figechoes3D}).
\par
\begin{figure}[ht] 
\centering 
\includegraphics[width=.8\columnwidth]{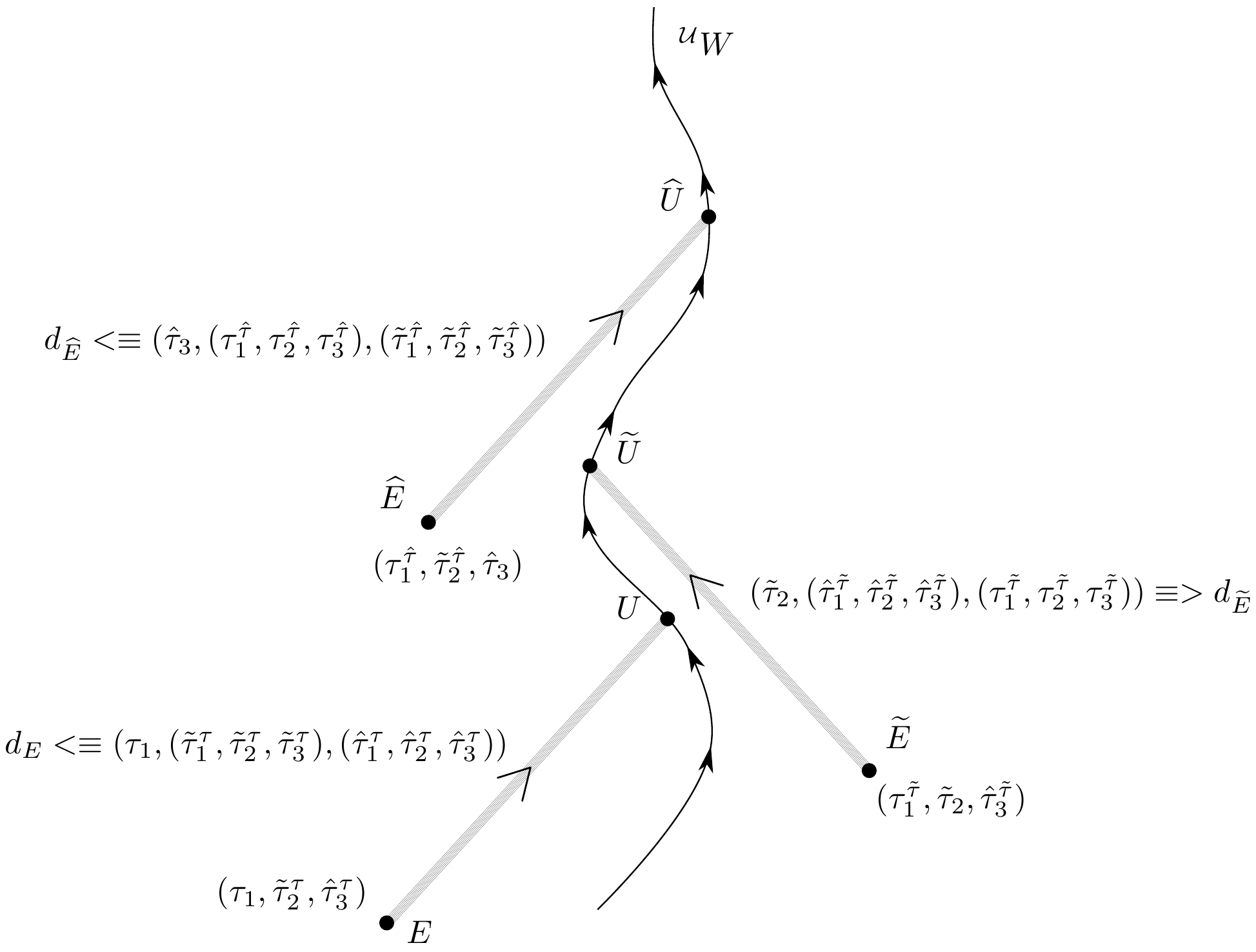}
\caption{\footnotesize\quad The sets of time stamps defining, with the three identifiers $id_\Em$, $id_\Emt$ and $id_\Emh$, the three data $d_E$, $d_{\Evt}$ and $d_\Evh$ received and recorded by the user $\Us$ at, respectively, $U$, $\widetilde U$ and $\widehat U$ on the worldline ${}^{\Us}W$ of $\Us$.}
\label{threedataU}
\end{figure}

Actually, the user receives $3\times7$ time stamps, \ie, three sets of data, namely, $d_E$, $d_\Evt$ and $d_\Evh$ such that
\begin{align*}
&d_E\equiv(\tau_1,(\taut^\tau_1,\taut^\tau_2,\taut^\tau_3),(\tauh^\tau_1,\tauh^\tau_2,\tauh^\tau_3), id_\Em)\quad\mbox{received at}\quad U\in {}^{\Us}W\,,\\
&d_\Evt\equiv(\taut_2,(\tauh^\taut_1,\tauh^\taut_2,\tauh^\taut_3),(\tau^\taut_1,\tau^\taut_2,\tau^\taut_3), id_\Emt)\quad\mbox{received at}\quad \widetilde{U}\in {}^{\Us}W\,,\\
&d_\Evh\equiv(\tauh_3,(\tau^\tauh_1,\tau^\tauh_2,\tau^\tauh_3),(\taut^\tauh_1,\taut^\tauh_2,\taut^\tauh_3), id_\Emh)\quad\mbox{received at}\quad \widehat{U}\in {}^{\Us}W\,.
\end{align*}
where $id_\Em$, $id_\Emt$ and $id_\Emh$ are identifiers of the emitters (see Fig.~\ref{threedataU}).
From now on, we consider only the sets of events represented in Fig.~\ref{figechoes3D}. 

\subsection{A first procedure of localization without stereometric grid}

The user can, then, also deduce three future light-like vectors generating the future null cone at $E$, namely, $\kh_E$, $\kt_E$ and $k^U_E$, such that
\[
\kh_E\equiv{}p_E-p_{\Evh'}\equiv\overrightarrow{E\Ph_E}\,,
\quad
\kt_E\equiv{}p_E-p_{\Evt'}\equiv\overrightarrow{E\Pt_E}\,,
\quad
k^U_E\equiv{}p_U-p_E\equiv\overrightarrow{EP^U_E}\,,
\]
where $P^U_E\equiv U$ and $p_U$ is the 3-position of $U$ in the emission grid. The three endpoints $\Ph_E$, $\Pt_E$ and $P^U_E$ define an affine plane $A_E$ in the emission grid. Then, a unique circumcircle in $A_E$ contains these three endpoints from which the unique circumcenter $C\in A_E$ can be deduced by standard formulas.\footnote{ 
\scriptsize
That is, we define the two relative vectors with origin $U$: $\tilde{r}=\kt_E-k^U_E$ and 
$\hat{r}=\kh_E-k^U_E$. Then, in $\Rset^3$, the circumcenter $C$ is the point $C\in A_E$ such that
\[
\overrightarrow{UC}=
\frac{(\|\tilde{r}\|^2\,\hat{r}-\|\hat{r}\|^2\,\tilde{r})\wedge(\tilde{r}\wedge\hat{r})}{2\,\|\tilde{r}\wedge\hat{r}\|^2}\,.
\]
\\ \parbox{3cm}{\hrulefill}
} 
\par
Now, let $e$ be an event to be localized in the emission grid (in this first procedure, the stereometric grid is not defined). It is featured and identified by a set $s_e$ of physical, non-geometrical characteristics such as, for instance, its spectrum, its shape, its temperature, etc. We assume also that this event $e$ can be detected and almost instantaneously physically analyzed by the emitters at the events $E$, $\Evt$ and $\Evh$ from signals carried by light rays (for instance) coming from $e$. Also, we assume that these light rays (which carry this various physical information) manifest themselves in ``bright points'' on their respective ``celestial circles'' of the emitters at the events $E$, $\Evt$ and $\Evh$. 
For the sake of illustration, we consider only the celestial circle $\Cscc\simeq S^1$ of the emitter $\Em$ at the event $E$. Also, we provide $\Em$ with an apparatus made of an optical device and a compass to locate the event $e$ on the celestial circle $\Cscc$.\footnote{
\scriptsize
The only remaining step utilizing material objects is the angle measurement by compasses. Their use implies that the angles remain invariant regardless of the size of the compass. And then, this also implies that there is an absolute notion of angle in contrast to the notions of time and length which depend on frames. This has historically been considered by Weyl and G{\"o}del with their concepts of  ``inertial compass'' or ``star compass'' in objection to Mach's principle. This absolute feature cannot come from any geometry of space-time. It is therefore possible that it comes from a different physics, such as quantum mechanics. Thus, a true compass would be based on the use of a quantum phenomenon of angle measurement, \ie, a \textit{quantum compass}.
This can be done with a Michelson interferometer (see for example \cite{E.Schrodinger:1954aa,S.R.Shepard:2009aa}) or interferences in cold atomic gases. Nevertheless, we think that the compass should be rather graduated by fractional numbers, for instance, such as those appearing in the fractional Hall effect.
\\\parbox{3cm}{\hrulefill}
} 
For this, we need also to define a projective frame for $\Cscc$. For this purpose, the two other satellites $\Emt$ and $\Emh$ manifest themselves in ``bright points'' on $\Cscc$ ascribed to the two events $\Evt'$ and $\Evh'$ in the past null cone of $E$. Then, the projective point $[0]_E\in\Cscc$ is ascribed to  $\Evt'$ and $\kt_E$, and the projective point $[\infty]_E\in\Cscc$ is ascribed to  $\Evh'$  and $\kh_E$:
\begin{eqnarray*}
\Evt'\longleftrightarrow&[0]_E&\longleftrightarrow\kt_E\,,\\
\Evh'\longleftrightarrow&[\infty]_E&\longleftrightarrow\kh_E\,.
\end{eqnarray*}
Then, we assume that $\Rset P^1\simeq\Cscc=S^1$.
Note that we cannot ascribe to $k^U_E$ and $U$ a projective point $[1]_E\in\Cscc$ since $U$ is in the future null cone of $E$, and thus, no corresponding ``bright points'' exists on $\Cscc$. Therefore, we need a fourth satellite, namely, $\EmS$, in addition to $\Em$, $\Emt$ and $\Emh$. \textit{A priori}, $\EmS$ does not need to broadcast a supplementary time stamp, but it must be clearly identified with an identifier $id_\EmS$. Then, another fourth ``bright point'' ascribed to the third projective point $[1]_E\in\Cscc$ is observable on $\Cscc$ due to $\EmS$ sending its identifier $id_\EmS$ from the event $S$ (see Fig.~\ref{figechoes3D}):
\[
S\longleftrightarrow [1]_E\,.
\]
Now, $e$ can be localized in the emission grid by applying the following procedure.
\par
From the ``bright points'' $[\infty]_E$, $[0]_E$ and $[1]_E$, and the optical device and compass embarked on $\Em$, the optical observation of $e$ on $\Cscc$ provides a projective point $[\tan\alpha]_E\in\Cscc$ with $\alpha$ clearly, numerically evaluated from the projective frame $\Frap_E\equiv\{[\infty]_E,[0]_E,[1]_E\}$.\footnote{%
In this definition of $[\tan\alpha]_E$, the angles $\alpha$ vary over the interval $[-\pi/2,\pi/2]$ of range $\pi$ on the celestial circle rather than over the usual interval $[0,2\pi]$ of range $2\pi$. Another viewpoint is to consider the ``angles'' to vary within the extended set of real numbers $\overline{\Rset}\equiv[+\infty,-\infty]$, and then, to write $[\alpha]_E$ with $\alpha\in\overline{\Rset}$ instead of $[\tan\alpha]_E$ with $\alpha\in[-\pi/2,\pi/2]$\,.
} %
Moreover, to $[\tan\alpha]_E$ there correspond two vectors $\vec{v}^{\,+}_E$ and 
$\vec{v}^{\,-}_E$ such that
\[
\vec{v}^{\,\pm}_E\equiv
\overrightarrow{EV^\pm_E}\equiv
\overrightarrow{EC}\pm
\left(
\overrightarrow{C\Pt_E}+\tan\alpha\,\overrightarrow{C\Ph_E}
\right),
\]
where $C$ is the circumcenter in $A_E$ and, in addition, $\overrightarrow{C\Pt_E}$ and $\overrightarrow{C\Ph_E}$ are ascribed to the following projective points:
\begin{eqnarray*}
\overrightarrow{C\Pt_E}&\longleftrightarrow&[0]_E\,,\\
\overrightarrow{C\Ph_E}&\longleftrightarrow&[\infty]_E\,.
\end{eqnarray*}
Now, the two vectors  $\vec{v}^{\,\pm}_E$ define a two dimensional affine plane 
$\Pcl_e$ containing $e$ such that
\[
\overrightarrow{Ee}=a^+\,\vec{v}^{\,+}_E+a^-\,\vec{v}^{\,-}_E\in\Pcl_e
\]
for two reals $a^\pm$ to be determined by applying the same procedure with the two emitters $\Emt$ and $\Emh$ at, respectively, $\Evt$ and $\Evh$. Indeed, we deduce the two other analogous affine planes $\Ptcl_e$ and $\Phcl_e$ and two relations as
\begin{align*}
&\overrightarrow{\Evt{e}}=\tilde{a}^+\,\vec{\tilde{v}}^{\,+}_E+\tilde{a}^-\,\vec{\tilde{v}}^{\,-}_E\in\Ptcl_e\,,\\
&\overrightarrow{\Evh{e}}=\hat{a}^+\,\vec{\hat{v}}^{\,+}_E+\hat{a}^-\,\vec{\hat{v}}^{\,-}_E\in\Phcl_e\,.
\end{align*}
Then, $e$ is the intersection point of $\Pcl$,  $\Ptcl_e$ and $\Phcl_e$.
Therefore, we obtain six algebraic linear equations determining completely the $a$'s and then $e$ in the emission grid. Neither stereometric coordinates nor, \textit{a fortiori}, a stereometric grid need to be defined. But, this procedure cannot be generalized to higher dimensional spacetime manifolds: it is specific to the three dimensional case. Indeed, the intersection point of three, two by two non-parallel planes always exists in $\Rset^3$ whereas four, two by two parallel, two-dimensional hyperplanes do not always have intersection points in $\Rset^4$.
\par\smallskip

\subsection{The intrinsic procedure of localization}

\begin{figure}[ht] 
\centering 
\def\svgwidth{\columnwidth} 
\includegraphics[width=\columnwidth]{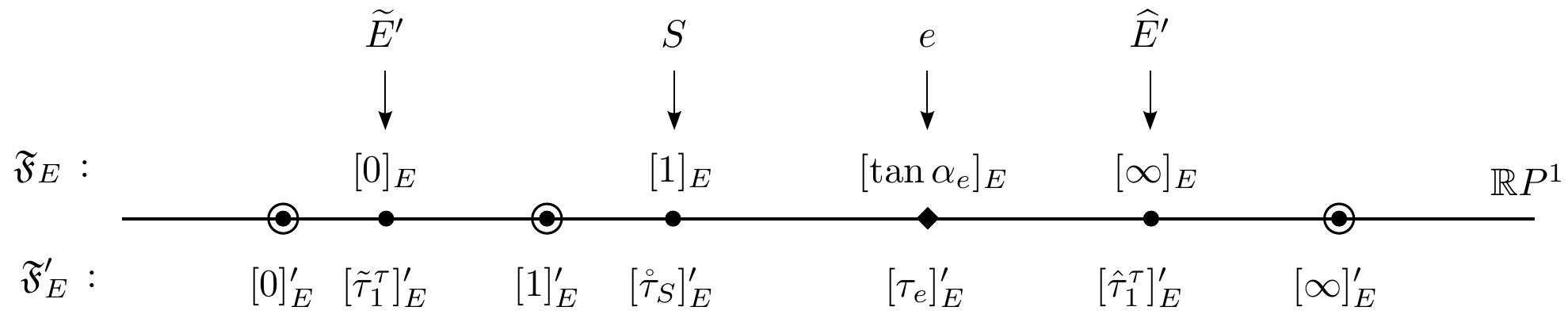}
\caption{\footnotesize\quad The change of projective frame at $E$.}
\label{chgtprojframE}
\end{figure}

A second, simpler, intrinsic and more effective procedure can be applied using again optical devices and compasses. It is based on a change of projective frame in $\Cscc$. More precisely, in the previous procedure with the projective frame $\Frap_E$ at $E$, the three projective points $[\infty]_E$, $[0]_E$ and $[1]_E$ defining $\Frap_E$ were ascribed to, respectively, $\Evh'$, $\Evt'$ and $S$. Now, we consider another projective frame $\Frap'_E\equiv\{[\infty]'_E,[0]'_E,[1]'_E\}$ such that
\begin{eqnarray*}
\Evt'&\longleftrightarrow&[\taut^\tau_1]'_E\,,\\
\Evh'&\longleftrightarrow&[\tauh^\tau_1]'_E\,,\\
S&\longleftrightarrow&[\taur_S]'_E\,,
\end{eqnarray*}
assuming now that $\EmS$ broadcasts also a fourth emission coordinate $\mathring{\tau}$ in addition to the three emission coordinates $\tau$, $\tilde\tau$ and $\hat\tau$. Then, in particular, $\EmS$ sends at the event $S$ the fourth time stamp $\mathring{\tau}_S$  received by $\Em$ at the event $E$  (see Fig.~\ref{figechoes3D}). Moreover, in a similar way, each other emitter $\Emt$ and $\Emh$ receives, respectively, at $\Evt$, the time stamp ${\mathring{\tau}}_{\widetilde{S}}$ and, at  $\Evh$, the time stamp ${\mathring{\tau}}_{\widehat{S}}$, from $\EmS$ at two events, respectively, $\widetilde{S}$ and $\widehat{S}$ in ${}^\EmS W$ differing in full generality from the event $S\in{}^\EmS W$. Hence, there are three corresponding emission events on the worldline of $\EmS$ for these three supplementary time stamps $\mathring{\tau}_S$, ${\mathring{\tau}}_{\widetilde{S}}$ and ${\mathring{\tau}}_{\widehat{S}}$. Then, there corresponds also to $e$ another projective point $[\tau_e]'_E$ with respect to this new projective frame $\Frap'_E$. As a consequence, the following correspondences 
\begin{eqnarray*}
{[0]_E}&\longleftrightarrow&[\taut^\tau_1]'_E\,,\\
{[\infty]_E}&\longleftrightarrow&[\tauh^\tau_1]'_E\,,\\
{[1]_E}&\longleftrightarrow&[\mathring{\tau}_S]'_E\,,\\
{[\tan\alpha_e]_E}&\longleftrightarrow&[\tau_e]'_E
\end{eqnarray*}
define the change of projective frame and, consequently, the projective point $[\tau_e]'_E$ (see Fig.~\ref{chgtprojframE}). 
\par
In homogeneous (projective) coordinates, this change of projective frame is defined by a matrix $K\in GL(2,\Rset)$ such that
\[
K\equiv
\begin{pmatrix}
a&b\\
c&d
\end{pmatrix},
\]
and satisfying the four following  additional correspondences:
\begin{eqnarray*}
{[0]_E}\equiv\begin{pmatrix}0\\1\end{pmatrix}&\xrightarrow{\quad K\quad}&\begin{pmatrix}a\\c\end{pmatrix}\equiv[\taut^\tau_1]'_E\quad\mbox{where $\taut^\tau_1=a/c$}\,,\\
{[\infty]_E}\equiv\begin{pmatrix}1\\0\end{pmatrix}&\xrightarrow{\quad K\quad}&\begin{pmatrix}b\\d\end{pmatrix}\equiv[\tauh^\tau_1]'_E\quad\mbox{where $\tauh^\tau_1=b/d$}\,,\\
{[1]_E}\equiv\begin{pmatrix}1\\1\end{pmatrix}&\xrightarrow{\quad K\quad}&\begin{pmatrix}a+b\\c+d\end{pmatrix}\equiv[\mathring{\tau}_S]'_E\quad\mbox{where}\quad\mathring{\tau}_S=\left(\frac{a+b}{c+d}\right),\\
{[\tan\alpha_e]_E}\equiv\begin{pmatrix}\tan\alpha_e\\1\end{pmatrix}&\xrightarrow{\quad K\quad}&
\begin{pmatrix}
a\,\tan\alpha_e+b\\
c\,\tan\alpha_e+d\end{pmatrix}
\equiv[\tau_e]'_E\,\,\mbox{where}\,\,
\tau_e=\left(\frac{a\,\tan\alpha_e+b}{c\,\tan\alpha_e+d}\right).
\end{eqnarray*}
Therefore, we obtain
\[
\begin{cases}
&a=-\,\tauh^\tau_1\,[\,\taut^\tau_1:\tauh^\tau_1:\mathring{\tau}_S\,]\,d\,,\\
&b=\taut^\tau_1\,d\,,\\
&c=[\,\taut^\tau_1:\tauh^\tau_1:\mathring{\tau}_S\,]\,d\,,
\end{cases}
\]
where $[\,\taut^\tau_1:\tauh^\tau_1:\mathring{\tau}_S\,]$ is such that
\[
[\,\taut^\tau_1:\tauh^\tau_1:\mathring{\tau}_S\,]\equiv
\left(
\frac{\taut^\tau_1-\mathring{\tau}_S}{\tauh^\tau_1-\mathring{\tau}_S}
\right).
\] 
Then, we deduce $\tau_e$ such that
\begin{equation}
\tau_e\equiv\left(\frac{\,\,\taut^\tau_1-\tauh^\tau_1\, [\,\taut^\tau_1:\tauh^\tau_1:\mathring{\tau}_S\,] \,\tan\alpha_e\,\,}{1- [\,\taut^\tau_1:\tauh^\tau_1:\mathring{\tau}_S\,]\,\tan\alpha_e }\right).
\label{transconftaue}
\end{equation}
This is a birational continuous function, and thus bijective.
In particular, we obtain the following important property: 
\begin{quote}
\textit{if $\tan\alpha_e=0$, $1$ or $\infty$, then we find $\tau_e=\taut^\tau_1$, $\mathring{\tau}_S$ or $\tauh^\tau_1$.}
\end{quote}
Also, from the other emitters at $\Evt$ and $\Evh$, the user can compute the three time stamps $p_e\equiv(\tau_e,\taut_e,\tauh_e)$ ascribed to the 3-position $p_e$ of the event $e$; therefore localized as expected.
However,  it is important to note that the event $e$ is not strictly located in the emission grid but in a new grid, namely, the \textit{stereometric grid}. More precisely, the stereometric grid shares with the emission grid 1) the three Cartesian time axes associated with the three emission coordinates $\tau$, $\tilde\tau$ and $\hat\tau$, and also, from the property above, 2) the three emitter worldlines only which are therefore common, point-to-point, to the two grids. Therefore, rigorously, if $e$ is not a point on an emitter worldline, then, $p_e$ must not be positioned (located) in the emission grid. Moreover, we begin the procedure with time stamps associated with events located in the emission grid, and we produce sets of time stamps to localize events. Then, because the process of location strictly differs from the process of localization, we must consider that any triplet $(\tau_e,\taut_e,\tauh_e)$ constitutes the three \textit{stereometric coordinates} ascribed to the event $e$ positioned in the stereometric grid only. This procedure can be outlined with the following diagram:
\[
\begin{CD}
\fbox{\parbox{2cm}{Emission\\ coordinates}}
@>>>
\fbox{\parbox{5cm}{Intrinsic angles\\ + non-intrinsic time stamps}}
@>>>
\fbox{\parbox{2.3cm}{Stereometric\\ coordinates}}
\end{CD}
\]
Also, it is important to note that given $E$, $\Evt$ and $\Evh$, the event $e$ is unique since it is the intersection point of three two-dimensional past null cones. Moreover, we can say that there exists a unique set of three events $E$, $\Evt$ and $\Evh$ ``attached'' to $e$, \ie, we have a fibered product of past null cones (over the set of localized events $e$ in $\Mmc$) homeomorphic to $\Mmc$. 
\par\smallskip
Hence, we need four satellites $\Em$, $\Emt$, $\Emh$ and $\EmS$ with their four emission coordinates to localize an event in the stereometric grid, and thus, the three dimensional spacetime $\Mmc$ must be embedded in $\Rset^4$. For instance, we have the following coordinates in $\Rset^4$:
\begin{subequations}
\begin{eqnarray}
{E}&\longleftrightarrow&(\tau_1,\taut^\tau_2,\tauh^\tau_3,\mathring{\tau}_S)\,,\\
\Evt&\longleftrightarrow&(\tau^\taut_1,\taut_2,\tauh^\taut_3,\mathring{\tau}_{\widetilde{S}})\,,\\
\Evh&\longleftrightarrow&(\tau^\tauh_1,\taut^\tauh_2,\tauh_3,\mathring{\tau}_{\widehat{S}})\,.
\end{eqnarray}
\label{4vectorsgrid}
\end{subequations}
Also, the data sent by the satellites $\Em$, $\Emt$ and $\Emh$ to the user $\Us$ are reduced. We just need the following reduced data:
\begin{align*}
&\bar{d}_E\equiv((\tau_1,\taut^\tau_2,\tauh^\tau_3,\mathring{\tau}_S), id_\Em,\alpha_e,s_e)\,,\\
&\bar{d}_\Evt\equiv((\tau^\taut_1,\taut_2,\tauh^\taut_3,\mathring{\tau}_{\widetilde{S}}), id_\Emt,\tilde{\alpha}_e, s_e)\,,\\
&\bar{d}_\Evh\equiv((\tau^\tauh_1,\taut^\tauh_2,\tauh_3,\mathring{\tau}_{\widehat{S}}), id_\Emh,\hat{\alpha}_e,s_e)\,,
\end{align*} 
where $s_e$ allows matching the three first data sets $d_E$, $d_{\Evt}$ and $d_{\Evh}$ ascribed to~$e$. 
\par\smallskip
Besides, the question arises to know if a fourth coordinate $\mathring{\tau}_{e}$ can be ascribed also to the event $e$ as for the three events $E$, $\Evt$ and $\Evh$. A coordinate $\mathring{\tau}_{e}$ could be easily obtained from the 3-position of $e$ in the stereometric grid if 1) $e$ is in the \textit{future horismos} \cite{kronpen67,garsenov05} of a point $p$ on the worldline of $\EmS$, and then, $\mathring{\tau}_{p}\equiv\mathring{\tau}_{e}$, and 2) $\EmS$ broadcasts also, in particular to the user,  the coordinates of $p$ in the stereometric grid obtained from the three other emitters $\Em$, $\Emt$ and $\Emh$. The first condition cannot always be physically or technologically satisfied since there necessarily exists an origin event $o$ at which the fourth satellite $\EmS$ begins to run. Hence, we can expect to know the positions of $\EmS$ in the stereometric grid only beyond this starting point $o$ on the future worldline ${}^{\EmS}W^+_o\equiv\{o\ll{p}, \mbox{\textit{where $p$ is an emission event of $\EmS$}}\}$ of $\EmS$ contained in the chronological future  of $o$ (the symbol $\ll$ denotes the \textit{chronological order}. See \cite{kronpen67,garsenov05} for instance). 
\par
Nevertheless, it is easy to circumvent this difficulty, assuming that we define the prolongation ${}^{\EmS}W^-_o$ of the worldline of $\EmS$ in the causal past of $o$ by a given, arbitrary, nevertheless well-defined by geometric conventions, curve in the stereometric grid. Now, from a given time parameterization of ${}^{\EmS}W^-_o$, we can also ascribe to any event $e$ a fourth time stamp $\mathring{\tau}_{e}$ from the message function $f^-_{{}^{\EmS}W^-_o}:e\longrightarrow\mathring{\tau}_{e}$ \cite{EhlPiSch:72}. Then, the worldline ${}^{\EmS}W$ of $\EmS$ is such that ${}^{\EmS}W={}^{\EmS}W^-_o\cup\{o\}\cup{}^{\EmS}W^+_o$ and we obtain the complete message function $f^-_{{}^{\EmS}W}:e\in\Mmc\longrightarrow\mathring{\tau}_{e}\in\Rset\simeq{}^{\EmS}W$\,. As a consequence, from $f^-_{{}^{\EmS}W}$\,, we obtain an embedding of $\Mmc$ in $\Rset^4$. This embedding is explicit since we cannot localize events without giving a fourth time stamp such as, for instance, ${\mathring{\tau}}_{\widetilde{S}}$.
\par
Furthermore, we recall that we have a local chart $\mu:([\tan\alpha_e],[\tan\tilde\alpha_e],[\tan\hat\alpha_e])\in(\Rset P^1)^3\longrightarrow{}p_e=(\tau_e,\taut_e,\tauh_e)\in\Rset^3$, and we consider now the action of $PGL(4,\Rset)$ on the triplets $(\tan\alpha_e,\tan\tilde\alpha_e,\tan\hat\alpha_e)$. Before, we denote by $\alpha_i$ ($i=1,2,3$) the three angles such that $\alpha_e\equiv\alpha_1$, $\tilde\alpha_e\equiv\alpha_2$ and $\hat\alpha_e\equiv\alpha_3$, and by $\tau_j$ ($j=1,2,3$) the three time stamps such that $\tau_e\equiv\tau_1$, $\taut_e\equiv\tau_2$ and $\tauh_e\equiv\tau_3$. We put below the list of formulas we start with. In particular,
we have a first set of formulas from the formulas such as \eqref{transconftaue} at $E\equiv E_1$, $\Evt\equiv E_2$ and $\Evh\equiv E_3$:
\begin{equation}
\tau_i=\left(\frac{
u_i^Q\,\tan\alpha_i+v_i^Q
}{
w_i^\ell\,\tan\alpha_i+k_i^\ell
}
\right)
\quad\text{at}\quad E_i\,,
\label{tauialphajqell}
\end{equation}
where we assume $w_i^\ell\neq0$ and where the superscripts {\scriptsize$Q$} and {\scriptsize$\ell$} indicate, respectively, that  $u_i^Q$, $v_i^Q$,
$w_i^\ell$ and $k_i^\ell$ are homogeneous polynomials of degree 2 ({\scriptsize$Q\equiv$} quadratic) and homogeneous polynomials of degree 1 ({\scriptsize$\ell\equiv$} linear) with respect to the set of time stamps collected at the three events $E_i$ for the localization of $e$. Also, we consider that any element $[P]\in PGL(4,\Rset)$ acts on the three tangents $\tan\alpha_i$ of the angles $\alpha_i$ to give three other tangents of angles $\tan\alpha'_j$ such that
\begin{equation}
\tan\alpha_i=\left(\frac{
\sum_{j=1}^3P_i^j\,\tan\alpha'_j+P_i^4
}{
\sum_{k=1}^3P_4^k\,\tan\alpha'_k+P_4^4
}
\right),
\label{alphaalphap}
\end{equation}
where $P\equiv(P^a_b)\in GL(4,\Rset)$ and $a,b=1,\dots,4$\,.
Then, replacing the three tangents $\tan\alpha_i$ in the formulas \eqref{tauialphajqell} by the three tangents $\tan\alpha_i$ given in the formulas \eqref{alphaalphap}, we obtain the following  second set of formulas:
\begin{equation}
\tau_i=\left(\frac{
\sum_{j=1}^3K_i^j\,\tan\alpha'_j+K_i^4
}{
\sum_{k=1}^3H_i^k\,\tan\alpha'_k+H_i^4
}
\right),
\label{taualphap}
\end{equation}
where the coefficients $K^a_b$ and $H^a_b$ ($a,b=1,\dotsc,4$) are linear with respect to the coefficients of $P\equiv(P^a_b)$. But, we can easily verify that these formulas can be rewritten in the following general form:
\begin{equation}
\tau_i=\left(\frac{
p_i^Q\,\tan\alpha'_i+q_i^Q
}{
r_i^\ell\,\tan\alpha'_i+s_i^\ell
}
\right),
\label{taualphapQell}
\end{equation}
which are of the same form as \eqref{tauialphajqell} where $p_i^Q$, $q_i^Q$, $r_i^\ell$ and $s_i^\ell$ depend on the remaining angles differing from $\alpha'_i$\,. In other words, any projective transformation $[P]$ provides admissible changes of projective frames from the given projective frames $\Frap_{E_i}$ to other projective frames $\Frap^\bullet_{E_i}$ on the celestial circles at the events $E_i$. These changes of projective frames are defined from the whole of the time stamps collected at the three events $E_i$ and not only at a given particular one. Thus, these changes differ from those from which we obtained, for instance, the formulas
\eqref{tauialphajqell}. As a consequence, the coefficients 
$p_i^Q$, $q_i^Q$, $r_i^\ell$ and $s_i^\ell$ depend on all of the time stamps and not only of those collected at the event $E_i$.
In addition, because we obtain admissible changes of projective frames, then any $[P]$ is an admissible projective transformation which can be, therefore, applied to the complete set of tangents, \textit{viz}, the set of tangents $([\tan\alpha'_1],[\tan\alpha'_2],[\tan\alpha'_3])$ in the present case or the set of tangents  $([\tan\alpha_1],[\tan\alpha_2],[\tan\alpha_3])$ as well.
\par
Besides, remarkably, the (non-unique) element $P\in GL(4,\Rset)$ such that,
for instance, 
\begin{subequations}
\begin{align}
&P^a_a=P^i_4=P^4_3=1\,,\qquad a=1,\dotsc4,\,\,i=1,2,3,\\
&P^4_1=P^3_1\,,\quad P^4_2=P^3_2\,,\\
&P^j_i=\frac{1}{w^\ell_i}\,(w_j^\ell+k_j^\ell-k_i^\ell)\,,\quad i\neq j\,,\,\,\,\,i,j=1,2,3,
\end{align}
\end{subequations}
gives the formulas \eqref{taualphap} with the same denominator for all the $\tau_i$, \ie, we have
\begin{equation}
\sum_{k=1}^3H_1^k\,\tan\alpha'_k+H_1^4
=\sum_{k=1}^3H_2^k\,\tan\alpha'_k+H_2^4
=\sum_{k=1}^3H_3^k\,\tan\alpha'_k+H_3^4\,.
\label{memedenomHalphap}
\end{equation}
More precisely, we obtain
\begin{equation}
H^k _i=w_k^\ell+k_k^\ell\,,
\qquad
H^4_i=w_3^\ell+k_3^\ell\,,
\end{equation}
for all $i,j=1,2,3$, and
\begin{equation}
K^a_i=\frac{1}{w^\ell_i}\,L^a_i
\end{equation}
for all $i=1,2,3$ and $a=1,\dotsc4$, where the $L$'s are homogeneous polynomials of degree 2 with respect to the coefficients $w_i^Q$, $u_i^Q$,
 $v_i^\ell$ and $k_i^\ell$. The element $P$ is not unique and we can obtain from other elements in $GL(4,\Rset)$ such a common denominator for the $\tau$'s.
\par\medskip
Beside, from this admissible definition of $P$, we define the \textit{virtual time stamps} $\tau^{vp}_i$ to be the limits obtained when the tangents $\tan\alpha'_i$ go to infinity. Then, we get stereometric points denoted by $\tau^{vp}$ which are ``aligned'' (element) in a two-dimensional affine subspace in the three-dimensional stereometric grid. We call  such points $\tau^{vp}$ \textit{vanishing points} or, equivalently,  \textit{points at infinity}. In addition, this subspace of vanishing points is (locally) homeomorphic to the two-dimensional projective real space $\Rset{P}^2$. It is important to note that any set of parallel infinite lines in the three-dimensional space (locally only homeomorphic to $\Rset{P}^3$) of the ``3-tangents'' $([\tan\alpha'_1],[\tan\alpha'_2],[\tan\alpha'_3])\in(\Rset{P}^1)^3$ are transformed by any $[P]$ into a congruence of infinite lines all crossing at a unique common vanishing point $\tau^{vp}$ in the stereometric grid. Hence, we can say, somehow, that each vanishing point is virtually ``spangled'' by a congruence of crossing lines defining  the extended notions of \textit{spatio-temporal perspective} or \textit{spatio-temporal parallax}. In addition, it is the so-called vanishing point of the projective geometry well-known by painters drawing perspectives on their canvas; hence the terminology. We suggest the existence of a sort of ``Big-Bang (visual) effect'' due to the ``spatio-temporal perspective'' relative to these vanishing points. We can note also, for example, that the particular event $\widehat{E}'$ can be possibly identified by localization with a vanishing point because one of its projective coordinates is $[\infty]_E$.

\subsection{Remarks and consequences}

From all these preliminary results, we can now deduce the following.
\begin{enumerate}
\item 
We have shown that any projective transformation $[P]\in PGL(4,\Rset)$
applied on the 3-tangents $([\tan\alpha_i])_{i=1,2,3}$ is compatible with changes of projective frames
 on the celestial circles of the three events, \textit{viz.}, $E$, $\widetilde{E}$ and $\widehat{E}$ (see Fig.~\ref{threedataU}), attached to any localized event $e$.
 \item
There always exists a particular projective transformation $[P]$ equalizing the denominators of the relations \eqref{taualphap} and such that these relations express another projective transformation (\textbf{PT}) in $PGL(4,\Rset)$ from the space of 3-tangents to the space of localized events. This has two consequences:
\begin{itemize}
\item The relations \eqref{taualphap} with the denominators equalized are the defining relations of a \textit{soldering map} from the projective space $\Rset P^3$ of 3-tangents to the stereometric grid of localized events in the spacetime manifold $\Mmc$. This soldering is a birational local map from $\Rset P^3$ to the stereometric grid of $\Mmc$. From \eqref{taualphap}, it is only a local map because
\begin{itemize}
\item[1)]  if the 3-tangent $\theta_e\equiv([\tan\alpha'_1],[\tan\alpha'_2],[\tan\alpha'_3])$ is considered as an element of $(\Rset{P}^1)^3$, \ie, $\theta_e\in(\Rset{P}^1)^3$ and $\theta_e$ goes to the unique limit  $\theta_{\infty}\equiv([\infty],[\infty],[\infty])$ in $(\Rset{P}^1)^3$, then, there corresponds to $\theta_{\infty}$ only one event $e_{\infty}\in\Mmc$, and, on the contrary,
\item[2)]  if $\theta_e$ is considered as an element of $\Rset{P}^3$, then $\theta_e$ has an infinite set of possible limits $\theta_{\infty}\equiv([\lambda],[\rho])\in\Rset{P}^2$.
\end{itemize}
Hence, assuming the soldering map to be non-local would involve 1) the wrong equivalence $(\Rset{P}^1)^3\simeq\Rset{P}^3$, and 2), that any direction 
$\theta_{\infty}\in\Rset{P}^2$ is completely identified with a unique corresponding spacetime event $e_{\infty}\in\Mmc$. Thus, we would go wrong in identifying a spacetime direction (\ie, a topological set of ``parallel'' lines in $\Mmc$) with a particular (unique) event in spacetime.
\par\smallskip
\item If $e^\star$ is another localized event attached to three other events $E^\star$, $\Evt^\star$ and $\Evh^\star$, then, there exists a \textbf{PT} from the stereometric coordinates $\tau_i^\star$ of $e^\star$ to the stereometric coordinates $\tau_i$ of $e$. Thus, $\Mmc$ is a so-called \textit{generalized Cartan space} ``modeled'' (locally) on the projective space $\Rset P^3$ (and not modeled on the vector space $\Rset^n$ usually associated  with any tangent vector space defined at every point of a differentiable manifold) \cite{Cartan:1925kx,Ehresm:50}.
\end{itemize}
\item
The \textbf{PT}s \eqref{taualphap} with \eqref{memedenomHalphap}
can be recast within the framework of the Lie groupoid structures.
For, we define, first, the \textit{data-point} $T_e$ to be the set of all of the time stamps collected at the events $E$, $\Evt$ and $\Evh$ to localize $e$, and, moreover, we denote by $\mathcal{T}$ the set of all such data-points $T_e$ as the localized event $e$ varies.
We assume $\mathcal{T}$ to be locally a smooth manifold. We shown that given two data-points $T_e$ and $T_{e^\star}$, then, the 3-position $p_{e^\star}$ is obtained from the 3-position $p_e$ by a \textbf{PT} defined explicitly
and univocally from $T_e$ and $T_{e^\star}$.
Hence, we can define the Lie groupoid $\mathcal{G}\rightrightarrows\mathcal{T}_s\times\mathcal{T}_t$ of \textbf{PT}s such that $\pi_s:\mathcal{G}\longrightarrow\mathcal{T}_s\equiv\mathcal{T}$ is the \textit{source map} and 
$\pi_t:\mathcal{G}\longrightarrow\mathcal{T}_t\equiv\mathcal{T}$ is the \textit{target map} of the groupoid. Then, the \textbf{PT}s deduced from any pair $(T_e,T_{e^\star})\in\mathcal{T}_s\times\mathcal{T}_t$ define sections of $\mathcal{G}$.
We can say that the translations from the source $T_e\in\mathcal{T}_s$ to the target $T_{e^\star}\in\mathcal{T}_t$ are in one-to-one correspondence with a 
\textbf{PT} defining $p_{e\star}$ from $p_e$. In other words, the projective structure given by this set of \textbf{PT}s is not, \textit{a priori}, strictly defined on $\Mmc$ but rather on the \textit{data manifold} $\mathcal{T}$. Nevertheless, to any data-point $T_e$ there corresponds a unique localized event $e$ relative to the given RPS. The reciprocal is less obvious but it is also true. Indeed, $e$ is the unique intersection point of three past null cones and only one triplet of such null cones have their apexes $E$, $\Evt$ and $\Evh$ on the worldlines of the three emitters $\Em$, $\Emt$ and $\Emh$. Therefore, once the worldlines of $\Em$, $\Emt$ and $\Emh$, $\EmS$ are known from this given RPS, then all the data needed to localize $e$ can be reached, and thus, $T_e$.
Hence, we can say also that we have a Lie groupoid structure on $\Mmc$ meaning that given $p_e$ and $p_{e^\star}$ only we can deduce the unique 
\textbf{PT} compatible with the localization process to pass from $p_e$ to $p_{e^\star}$. This \textbf{PT} is not applied to the whole of the events in the stereometric grid. It is not a \textbf{PT} of the stereometric grid.
\par
Also, we can say that a mere translation from $p_e$ to $p_{e^\star}$ in the stereometric grid is, somehow, ``converted'' to a \textbf{PT} ``compatible'' with the localization process.
By ``compatible,'' we mean that the translations, for instance, in the stereometric grid cannot be directly and physically observed by the use of an explicit causal protocol, unlike the admissible \textbf{PT}s on the celestial circles. And, moreover, assuming that we are not permanently drunk, ``lucidly'' looking at two simultaneous realities hierarchized according to our degree of consciousness into an ``appearance'' and a ``reality,'' then, if we see only one ``manifest image'' \cite{Rosset:2012aa,frassen99,Rosset:1993ab} on each celestial circle, then, this is just ``the'' reality...
Thus, those transformations, such as the translations or any transformation in the affine group, must be interpreted or, somehow, ``converted'' into a manifest
\textbf{PT}. But, we can avoid such conversion or interpretation considering that the grid has the structure of a projective space onto which transformations in the affine group, for instance, are forbidden, useless or not physical because physically not manifest or obervable via a causal protocol.
\par
From a more mathematical viewpoint, if, on the one hand, the (finite) local \textbf{PT}s are defined as elements of a Lie groupoid $\mathcal{G}$ over $\Mmc\times\Mmc$, then, on the other hand, from the present particular groupoid structure, the corresponding Lie algebroid is just identified with the module of vector fields on $\Mmc$. In other words, the tensorial calculus must be a projective tensorial calculus over $\Mmc$. As a consequence, the connections on $\Mmc$ must be projective Cartan connections.
\par
Moreover, the latter can be restricted to reduced projective connections on each celestial circle in accordance with a mathematical procedure/com\-pu\-ta\-tion analogous to the one giving the transformation formulas \eqref{taualphapQell} on each celestial circle from the general transformation formulas \eqref{taualphap} on the whole of $\Mmc$.
\par
Hence, because the data space $\mathcal{T}$
is locally homeomorphic to $\Mmc$ (we assume that it is, actually, diffeomorphic), we can make the geometrical computations on $\Mmc$ in an abstract way, \ie, avoid considering the full set of time stamps of $T_e$ and considering only the restricted set of time stamps directly identified with $p_e$ as much as only infinitesimal, tensorial computations are carried out; and thus, the origin of the ``infinitesimal'' projective geometry of $\Mmc$ (but the finite projective geometry on $\Mmc\times\Mmc$ via the groupoid $\mathcal G$).
\end{enumerate}

Lastly, we call the worldline ${}^{\EmS}W$ of the emitter $\EmS$ an \textit{anchoring worldline}, and we call the event $a\in{}^{\EmS}W$ such that the time stamp $\taur_{a}$ emitted by $\EmS$ at $a$ is such that $\taur_{a}=f^-_{{}^{\EmS}W}(e)$ and $\taur_{a}\equiv\taur_{e}$ the \textit{anchor} $a$ of $e$.

\section{The protocol of localization in a $(3+1)$-dimensional spacetime $\Mmc$ modeled on $\Rset P^4$  \label{sec4D}}

The generalization of the previous protocol follows a similar process with five emitters $\Em$, $\Emb$, $\Emt$, $\Emh$ and $\Emr$ associated with five emission coordinates, respectively, $\tau$, $\bar\tau$, $\tilde\tau$, $\hat\tau$ and $\taur$\,.
They constitute five RPSs made up, each, of four emitters among these five with the fifth one used for the localization of spacetime events denoted by $e$. Also, as in the preceding sections, we denote the user by $\Us$  and the celestial spheres of the five emitters by, respectively, $\Ccl$, $\Cclb$, $\Cclt$, $\Cclh$ and $\Cclr$. The five emission grids of these five RPSs are Euclidean spaces $\Rset^4$. The passage from any emission grid to another one among the four others is a change of chart which is well-defined once the dated trajectories of the five emitters in the grids are obtained from each RPS and recorded. 
\par\smallskip
For the sake of argument, we consider only the RPS made with the first four emitters, namely,  $\Em$, $\Emb$, $\Emt$ and $\Emh$ and its associated emission grid with the four time stamps  $\tau$, $\bar\tau$, $\tilde\tau$ and $\hat\tau$ defining the so-called 4-positions of the events in this emission grid. Then, the fifth emitter $\Emr\equiv\EmS$ is used to complement this, for the localization process. Consequently, the worldline $\mathring{W}$ of $\Emr$ is the anchoring worldline of the relativistic localization system.
\par
\begin{figure}[ht] 
\centering 
\includegraphics[width=.8\columnwidth]{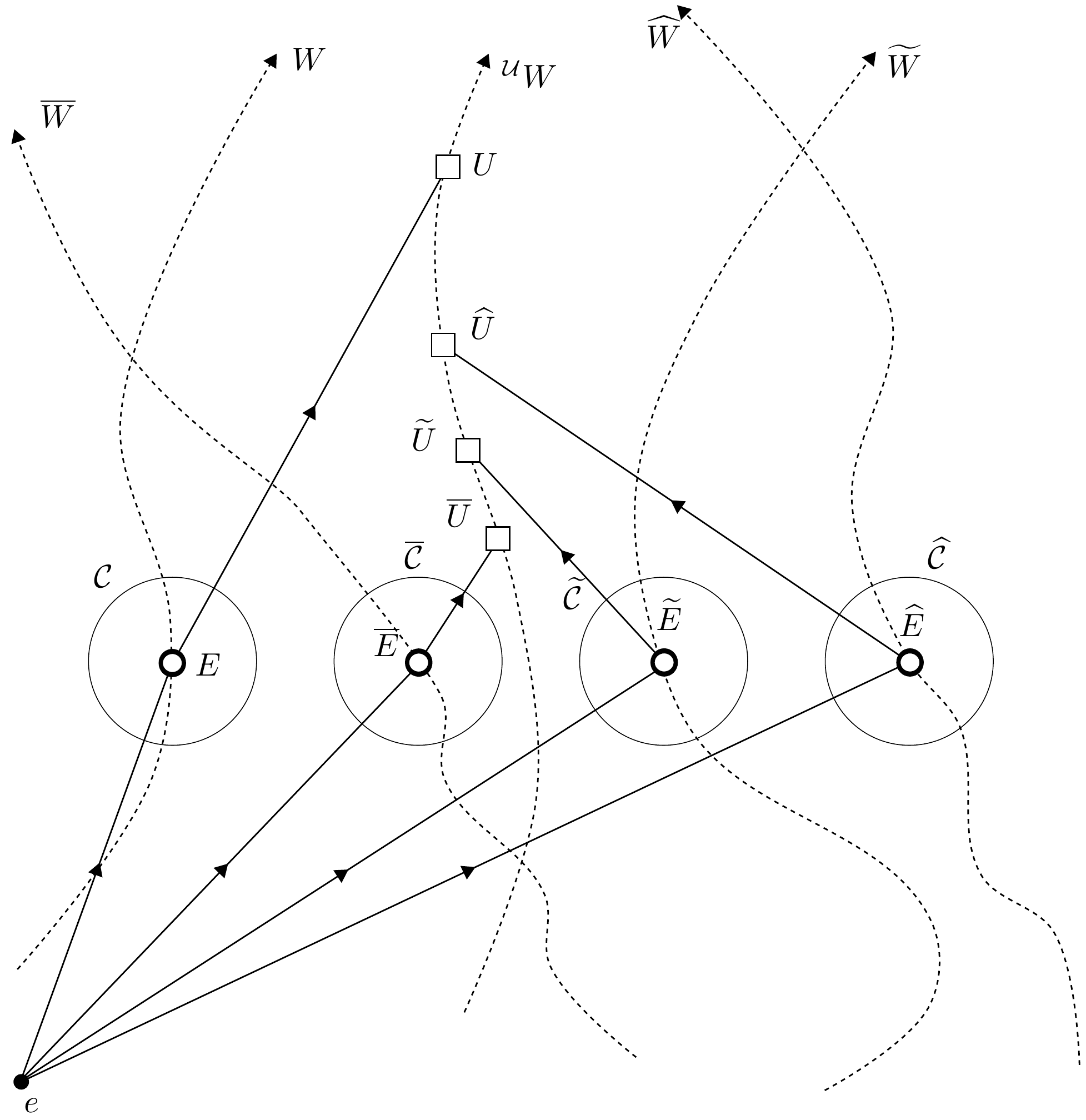}
\caption{\footnotesize\quad The event $e$ in the four past  null cones of the four events $E$, $\Evb$, $\Evt$ and $\Evh$. This event $e$ is observed on their respective celestial spheres $\Ccl$, $\Cclb$, $\Cclt$ and $\Cclh$.}
\label{EventsatU}
\end{figure}

\begin{figure}[ht] 
\centering 
\includegraphics[width=.85\columnwidth]{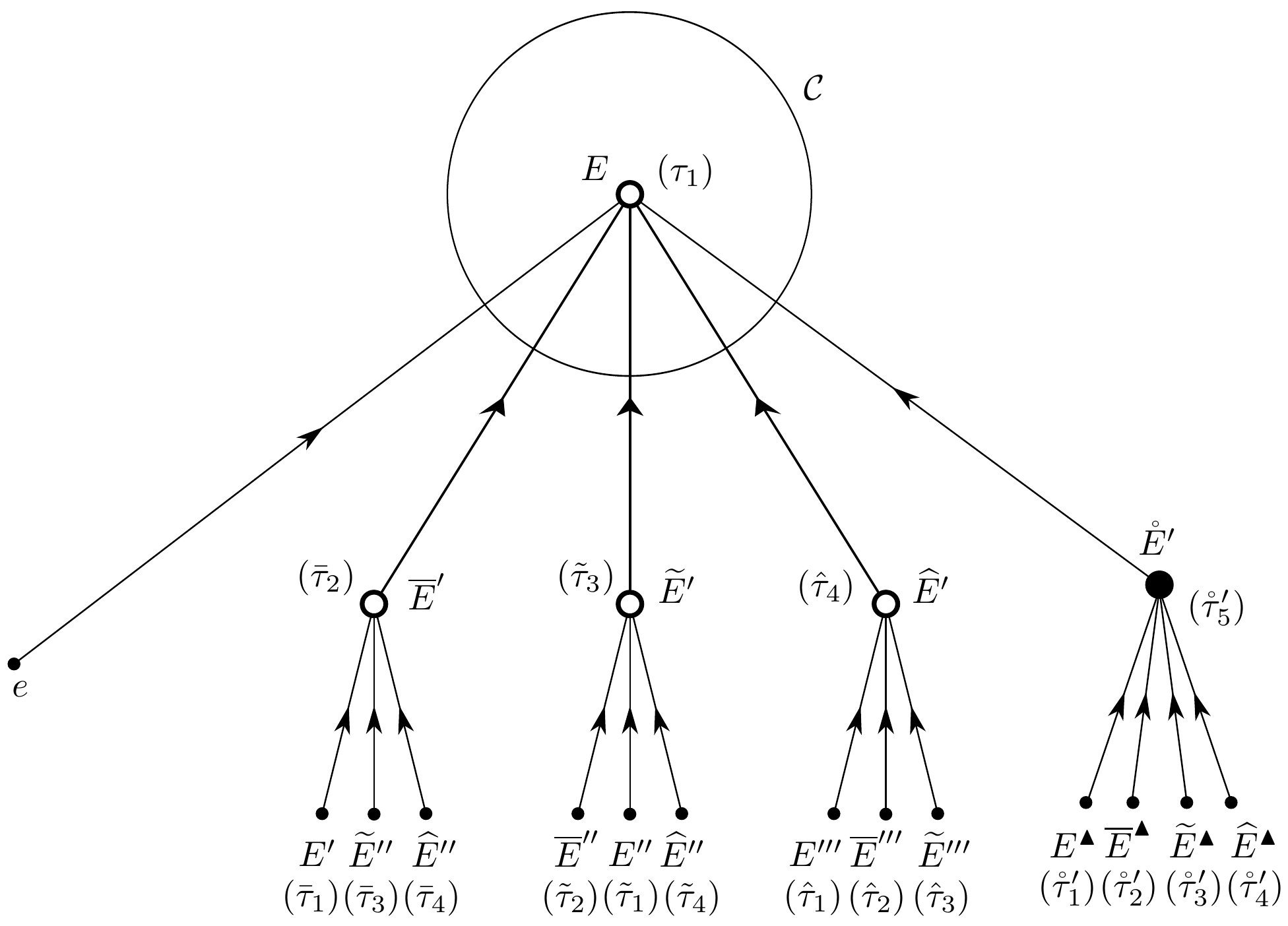}
\caption{\footnotesize\quad 
The event $E$ in the five future  null cones of the five events $e$, 
$\Evb'$, $\Evt'$, $\Evh'$ and $\Evr'$.}
\label{EventsatE}
\end{figure}

\begin{figure}[ht] 
\centering 
\includegraphics[width=.55\columnwidth]{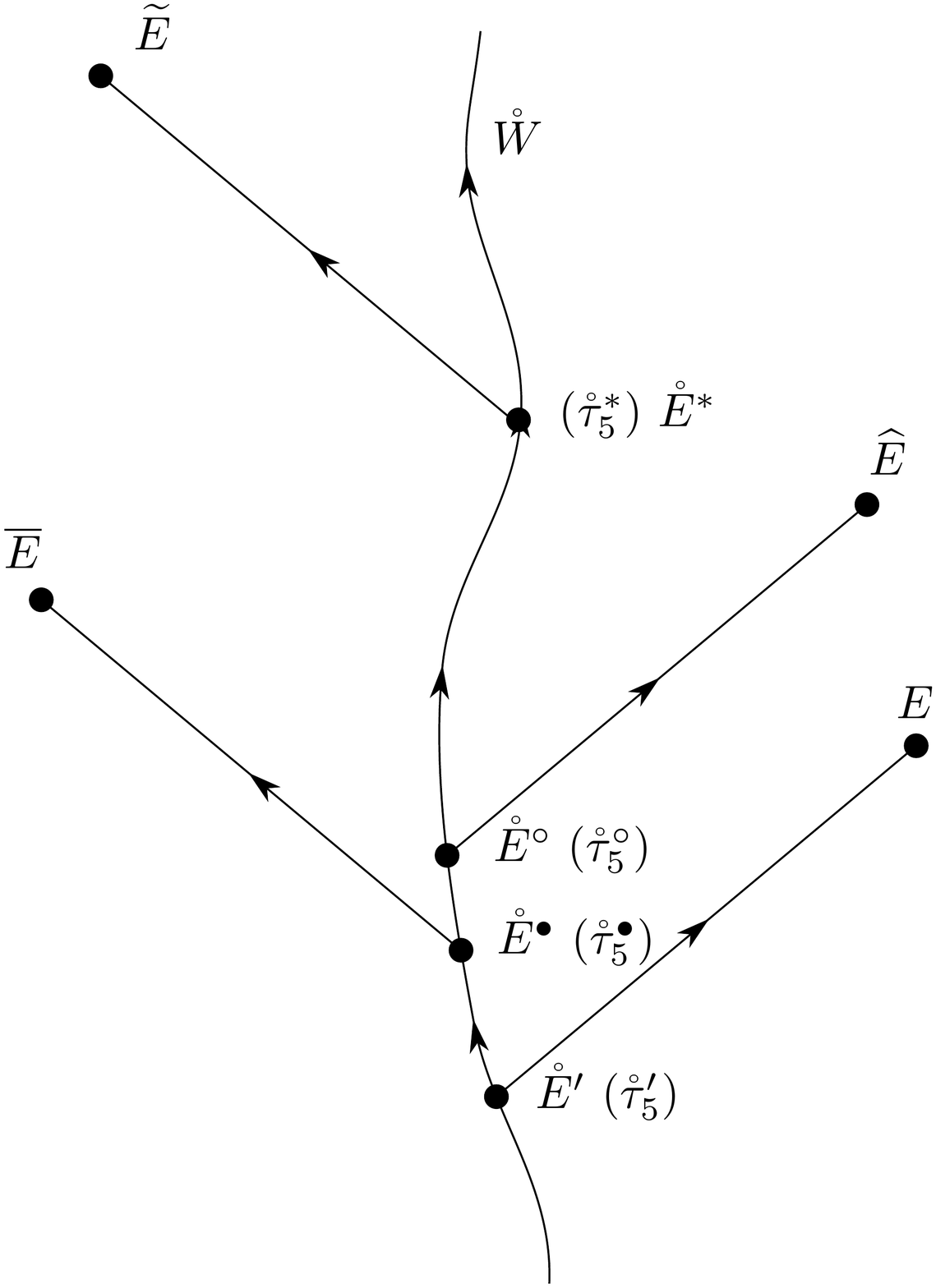}
\caption{\footnotesize\quad An example of successive events $\Evr'$, $\Evr^\bullet$, $\Evr^\circ$ and $\Evr^*$ on the anchoring worldline of $\Emr$ transmiting their coordinates $\taur_5$ towards the four events $E$, $\Evb$, $\Evh$ and  $\Evt$.}
\label{treeevnts}
\end{figure}

\smallskip
Now,  we consider only the set of particular events represented in Figs.~\ref{EventsatU}, \ref{EventsatE} and \ref{treeevnts} with their corresponding tables of 4-positions. 
\par
Fig.~\ref{EventsatU} shows the different events, namely, $E$ on the worldline $W$ of $\Em$, $\Evb$ on the worldline $\overline{W}$ of $\Emb$, $\Evt$ on the worldline $\widetilde{W}$ of $\Emt$ and $\Evh$ on the worldline $\widehat{W}$ of $\Emh$, at which the event $e$ is manifest on their respective celestial spheres.
We assume that the data of localization for $e$ collected at the events $E$, $\Evb$, $\Evt$ and $\Evh$ are sent to the user and they are received at the events, respectively, $U$, $\overline{U}$, $\widetilde U$ and $\widehat U$ on the worldline ${}^\Us W$ of $\Us$.
\par
Fig.~\ref{EventsatE} indicates, first, the events $\Evb'$, $\Evt'$ and $\Evh'$ from which the 4-position  of the event $E$ can be known in the emission grid (see Table \ref{table4pos}) and, second, two other events, namely, $\Evr'$ and $e$, which are observed on the celestial sphere $\Ccl$ of the emitter $\Em$ at $E$. Obviously, $e$ is the event to be localized and $\Evr'$ is a particular event on the worldline of $\Emr$ which broadcasts the time stamp $\taur'_5$ to $E$ used for the localization process. Similar figures could be indicated concerning the three other events $\Evb$, $\Evt$ and $\Evh$ on Fig.~\ref{EventsatU}, but they are not really necessary for the description of the localization process presented below. These unnecessary supplementary figures would indicate supplementary events on the worldline of $\Emr$, such as, for instance, $\Evr^\bullet$ from which (see Fig.~\ref{treeevnts}) the time stamp $\taur^\bullet_5$ is transmitted to the event $\Evb$ of Fig.~\ref{EventsatE}.
These particular time stamps are denoted by $\taur_5$ (with different superscripts) and they are sent from different events on the worldline of $\Emr$  to the other four emitters.

\begin{table}[htdp]
\begin{center}
\begin{tabular}{|c|c|}
\hline\hline
\rule[-2ex]{0ex}{5ex}Event&4-position\\
\hline\hline
\rule[-1.5ex]{0ex}{4.5ex}$\Evb'$&$(\taub_1,\taub_2,\taub_3,\taub_4)$\\ \hline
\rule[-1.5ex]{0ex}{4.5ex}$\Evt'$&$(\taut_1,\taut_2,\taut_3,\taut_4)$\\ \hline
\rule[-1.5ex]{0ex}{4.5ex}$\Evh'$&$(\tauh_1,\tauh_2,\tauh_3,\tauh_4)$\\ \hline
\rule[-1.5ex]{0ex}{4.5ex}$E$&$(\tau_1,\taub_2,\taut_3,\tauh_4)$\\ \hline
\rule[-1.5ex]{0ex}{4.5ex}$\Evr'$&$(\taur'_1,\taur'_2,\taur'_3,\taur'_4)$\\ 
\hline\hline
\end{tabular}
\caption{\footnotesize\quad The 4-positions of the events in Fig.~\ref{EventsatE}.\label{table4pos}}
\end{center}
\end{table}
\begin{figure}[ht] 
\centering  
\includegraphics[width=.7\columnwidth]{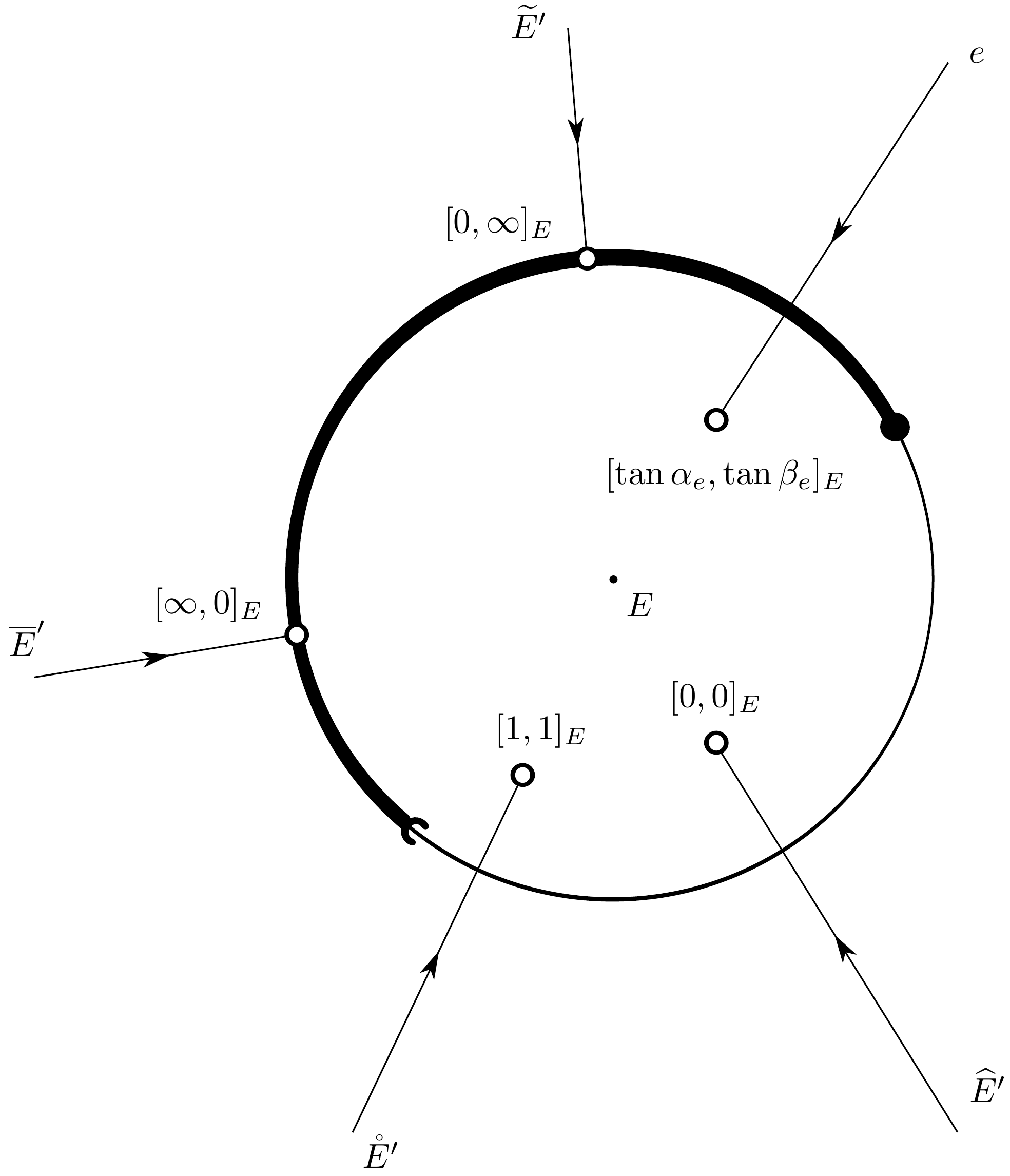}
\caption{\footnotesize\quad The projective disk  at the event $E$ associated to the celestial sphere $\Ccl$ of the emitter $\Em$.}
\label{projdisk}
\end{figure}
Also, angles are evaluated on each celestial sphere from optical devices and compasses providing pairs of angles, namely, $(\alpha,\beta)$ ascribed to each ``bright point'' observed and tracked on any given celestial sphere. 
Actually, each celestial sphere (homeomorphic to $S^2$) is considered as the union of a circle and two hemispheres. They are topological sets of which the first one is a closed set and also the common boundary of the others, which are two open sets in $S^2$. In addition, each hemisphere is embedded in an open, connected and simply connected set in $\Rset P^2$ and, moreover, each hemisphere is supplied with a given projective frame made of four particular points to be specified in the sequel.
\par
One hemisphere is made of a little spherical cap, as small as possible, and the other is its complementary hemisphere in $S^2$ with their common boundary to be, for instance, a polar circle. This choice is motivated from metrological considerations. Indeed, we want the probability of passage from one hemisphere to the other to be as small as possible when tracking trajectories of moving points on the celestial spheres. Nevertheless, we provide each celestial sphere with a  computing device ensuring, on the polar circle, the change of projective frame  from one hemisphere to the other and, for each moving point, recording the signature of its passage, \textit{viz}, a plus or minus sign. As a consequence, we can track more completely moving  ``bright points,'' and then, we can position these points in only one specified, given system of projective coordinates common to the two hemispheres minus a point (the north pole for instance) to which is ascribed an identifying symbol instead of two angles. Then, we can establish the correspondences between the pairs of angles in the two hemispheres and on the polar circle. 
\par
We usually represent one hemisphere embedded in $\Rset P^2$ by a two-dimen\-sional disk in $\Rset^2$ to which is added one-half of the polar circle. Then, we have projective frames made of the four projective points $[\infty,0]$, $[0,\infty]$, $[0,0]$ and $[1,1]$ with the first two on the polar circle (see Fig.~\ref{projdisk}). Also, a projective point $[\tan\alpha_e,\tan\beta_e]$ is ascribed to the event $e$ observed on each celestial sphere.
More precisely, one of the two projective spaces $\Rset P^2$ attached to the celestial sphere $\Ccl$ of $\Em$ at the event $E$ is represented in Fig.~\ref{projdisk}. Also, a  first projective frame $\Frap_E\equiv\{[\infty,0]_E,[0,\infty]_E,[0,0]_E,[1,1]_E\}$ attached to this projective space is represented providing the projective coordinates  $[\tan\alpha,\tan\beta]_E$. Also, a second projective frame $\Frap'_E\equiv\{[\infty,0]'_E,[0,\infty]'_E,[0,0]'_E,[1,1]'_E\}$ is defined from a change of projective frame from $\Frap_E$ to  $\Frap'_E$. This change of frame is based on pairs of numerical values given, for instance, by the first pair of time stamps, namely, $(\tau_1,\tau_2)$ obtained from the first emitters $\Em$ and $\Emb$. 
\par\bigskip
More precisely, we define the first four correspondences:
\[
\begin{matrix}
e&\longleftrightarrow&[\tan\alpha_e,\tan\beta_e]_E&\longleftrightarrow&[\tau^E_e,\taub^E_e]'_E\,,\\
\Evb'&\longleftrightarrow&[\infty,0]_E&\longleftrightarrow&[\taub_1,\taub_2]'_E\,,\\
\Evt'&\longleftrightarrow&[0,\infty]_E&\longleftrightarrow&[\taut_1,\taut_2]'_E\,,\\
\Evh'&\longleftrightarrow&[0,0]_E&\longleftrightarrow&[\tauh_1,\tauh_2]'_E\,,
\end{matrix}
\]
but with the additional correspondence
\[
\begin{matrix}
\Evr'&\longleftrightarrow&[1,1]_E&\longleftrightarrow&[\taur'_5,\lambda]'_E
\end{matrix}\,,
\]
where $\lambda$ is a time value free to vary at this step of the process.
Other correspondences can be chosen. All can be brought back to any fixed, given one once the changes of charts between the five possible emission grids are known. Thus, one correspondence only can be used to present the localization protocol.
\par
Also, it is important to note that $\taur'_5$ can be one of the four other time stamps received at $\Evr'$ by $\Emr$ from the four other satellites, \ie, it can be equal to $\taur'_1$, $\taur'_2$, $\taur'_3$ or $\taur'_4$. But, these four values are clearly independent on the whole of the other time stamps such as, for instance, $\tau_1$, $\tauh_3$, $\taut_4$, etc., involved in the localization process, all the more so since  these time stamps $\taur'_i$ depend on the worldline of $\Emr$. Hence, $\taur'_5$ is considered as  an independent time variable in the process---so, a fifth supplementary time stamp indexed by the number 5.
In addition, the parameter $\lambda$ is, actually, well-defined, as shown in the sequel, from the complete description of the process of localization.
\par
Furthermore, we can set the Table~\ref{tableattrib} of attributions based on the following pairs of time stamps: $\tau_1$ and $\tau_2$ for $E$,
$\tau_2$ and $\tau_3$ for $\Evb$,  $\tau_3$ and $\tau_4$ for $\Evt$, and
 $\tau_4$ and $\tau_1$ for $\Evh$ (only the correspondences [angles] $\longleftrightarrow$ [time stamps] are indicated in this table; the others are not need for the explanations given below and they are indicated by the marks ``$***$''): 
\begin{table}[htdp]
\caption{Attributions of time stamps, angles and events.}
\begin{center}
\begin{tabular}{||c||c|c|c|c|c|c|c||}
\hline\hline
\rule[-1em]{0ex}{2.5em}\makebox[2em]{}&\makebox[3em]{$\Em$}&\makebox[3em]{$\Emb$}&\makebox[3em]{$\Emt$}&\makebox[3em]{$\Emh$}&\makebox[3em]{$\Emr$}&\makebox[3em]{event}&\parbox{5.2em}{\raisebox{-.2em}{pair of}\\ \raisebox{.2em}{time stamps}}\\
\hline\hline
\rule[-.5em]{0ex}{2em}&---&$\Evb'$&$\Evt'$&$\Evh'$&$\Evr'$&$e$&\\ \cline{2-7}
\rule[-.75em]{0ex}{2.25em}$E$&---&$[\infty,0]_E$&$[0,\infty]_E$&$[0,0]_E$&$[1,1]_E$&$[\tan\alpha_e,\tan\beta_e]_E$&$(\tau_1,\tau_2)$\\ \cline{2-7}
\rule[-.75em]{0ex}{2.25em}&---&$[\taub_1,\taub_2]'_E$&$[\taut_1,\taut_2]'_E$&$[\tauh_1,\tauh_2]'_E$&$[\taur'_5,\lambda]'_E$&$[\tau^E_e,\taub^E_e]'_E$&\\ \hline\hline
\rule[-.5em]{0ex}{2em}&$E^\bullet$&---&$\Evt^\bullet$&$\Evh^\bullet$&$\Evr^\bullet$&$e$&\\ \cline{2-7}
\rule[-.75em]{0ex}{2em}$\Evb$&$[\infty,0]_{\Evb}$&---&$[0,0]_{\Evb}$&$[0,\infty]_{\Evb}$&$[1,1]_{\Evb}$&$[\tan\bar\alpha_e,\tan\bar\beta_e]_{\Evb}$&$(\tau_2,\tau_3)$\\ \cline{2-7}
\rule[-.75em]{0ex}{2em}&***&---&***&***&$[\taur^\bullet_5,\bar\lambda]'_{\Evb}$&$[\taub^{\Evb}_e,\taut^{\Evb}_e]'_{\Evb}$&\\ 
\hline\hline
\rule[-.5em]{0ex}{2em}&$E^*$&$\Evb^*$&---&$\Evh^*$&$\Evr^*$&$e$&\\ \cline{2-7}
\rule[-.75em]{0ex}{2em}$\Evt$&$[0,\infty]_{\Evt}$&$[0,0]_{\Evt}$&---&$[\infty,0]_{\Evt}$&$[1,1]_{\Evt}$&$[\tan\tilde\alpha_e,\tan\tilde\beta_e]_{\Evt}$&$(\tau_3,\tau_4)$\\ \cline{2-7}
\rule[-.75em]{0ex}{2em}&***&***&---&***&$[\taur^*_5,\tilde\lambda]'_{\Evt}$&$[\taut^\Evt_e,\tauh^\Evt_e]'_{\Evt}$&\\ 
\hline\hline
\rule[-.5em]{0ex}{2em}&$E^\circ$&$\Evb^\circ$&$\Evt^\circ$&---&$\Evr^\circ$&$e$&\\ \cline{2-7}
\rule[-.75em]{0ex}{2em}$\Evh$&$[0,0]_{\Evh}$&$[0,\infty]_{\Evh}$&$[\infty,0]_{\Evh}$&---&$[1,1]_{\Evh}$&$[\tan\hat\alpha_e,\tan\hat\beta_e]_{\Evh}$&$(\tau_4,\tau_1)$\\ \cline{2-7}
\rule[-.75em]{0ex}{2em}&***&***&***&---&$[\taur^\circ_5,\hat\lambda]'_{\Evh}$&$[\tauh^\Evh_e,\tau^\Evh_e]'_{\Evh}$&\\ 
\hline\hline
\end{tabular}
\end{center}
\label{tableattrib}
\end{table}%
\par
Then, we determine the change of projective frame in $\Rset P^2$ on the celestial sphere $\Ccl$ of $\Em$ at $E$. For this, we must compute the matrix $K$ as 
\begin{equation}
K=
\begin{pmatrix}
a&d&g\\
b&e&h\\
c&f&k
\end{pmatrix}
\end{equation}
associated with this change of frame. This matrix $K$ is defined from the following correspondences in $\Rset^3$:
\begin{align*}
\Evb'\,:\,\,&
[\infty,0]_E\equiv\begin{pmatrix}1\\0\\0\end{pmatrix}
\qquad
\xrightarrow{\quad K\quad}&\quad
[\taub_1,\taub_2]'_E\equiv\begin{pmatrix}a\\b\\c\end{pmatrix}
\quad\mbox{where}\quad
\begin{cases}
\taub_1&=a/c\\
\taub_2&=b/c
\end{cases}
\end{align*}
\begin{align*}
\Evt'\,:\,\,&
[0,\infty]_E\equiv\begin{pmatrix}0\\1\\0\end{pmatrix}
\qquad
\xrightarrow{\quad K\quad}&\quad
[\taut_1,\taut_2]'_E\equiv\begin{pmatrix}d\\e\\f\end{pmatrix}
\quad\mbox{where}\quad
\begin{cases}
\taut_1&=d/f\\
\taut_2&=e/f
\end{cases}
\end{align*}
\begin{align*}
\Evh'\,:\,\,&
[0,0]_E\equiv\begin{pmatrix}0\\0\\1\end{pmatrix}
\qquad
\xrightarrow{\quad K\quad}&\quad
[\tauh_1,\tauh_2]'_E\equiv\begin{pmatrix}g\\h\\k\end{pmatrix}
\quad\mbox{where}\quad
\begin{cases}
\tauh_1&=g/k\\
\tauh_2&=h/k
\end{cases}
\end{align*}
\begin{align*}
\Evr'\,:\,\,&
[1,1]_E\equiv\begin{pmatrix}1\\1\\1\end{pmatrix}
\qquad
\xrightarrow{\quad K\quad}&\,\,\,\,
[\taur'_5,\lambda]'_E\equiv\begin{pmatrix}a+d+g\\b+e+h\\c+f+k\end{pmatrix}
\,\mbox{where}\,
\begin{cases}
\taur'_5&=\left(\frac{a+d+g}{c+f+k}\right)\\
\lambda&=\left(\frac{b+e+h}{c+f+k}\right)
\end{cases}
\end{align*}
\begin{align*}
e\,:\,\,&
[\tan\alpha_e,\tan\beta_e]_E\equiv\begin{pmatrix}\tan\alpha_e\\\tan\beta_e\\1\end{pmatrix}
\xrightarrow{\quad K\quad}&
[\tau^E_e,\taub^E_e]'_E\equiv\begin{pmatrix}u\\v\\w\end{pmatrix}
\,\,\,\mbox{where}\,\,\,
\begin{cases}
\tau^E_e&=u/w\\
\taub^E_e&=v/w
\end{cases}
\end{align*}
and
\begin{align*}
u&=a\,\tan\alpha_e+d\,\tan\beta_e+g\,,\\
v&=b\,\tan\alpha_e+e\,\tan\beta_e+h\,,\\
w&=c\,\tan\alpha_e+f\,\tan\beta_e+k\,.
\end{align*}
From the above, we deduce the four following linear equations:
\begin{subequations}
\begin{align}
&
	\begin{cases}
	&(\taub_1-\taur'_5)\,x+(\taut_1-\tau'_5)\,y+(\tauh_1-\tau'_5)=0\,,\\
	&(\taub_2-\lambda)\,x+(\taut_2-\lambda)\,y+(\tauh_2-\lambda)=0\,,
	\end{cases}
	\label{Eqxy1}
\\
&
	\begin{cases}
	&(\taub_1-\tau^E_e)\,x\,\tan\alpha_e+(\taut_1-\tau^E_e)\,y\,\tan\beta_e+(\tauh_1-\tau^E_e)=0\,,\\
	&(\taub_2-\taub^E_e)\,x\,\tan\alpha_e+(\taut_2-\taub^E_e)\,y\,\tan\beta_e+(\tauh_2-\taub^E_e)=0\,,
	\end{cases}
	\label{tauEetaubEe}
\end{align}
\end{subequations}
where $x\equiv c/k$ and $y\equiv f/k$, and where $x$, $y$, $\lambda$, $\tau^E_e$ and $\taub^E_e$ are the unknowns. From the system \eqref{Eqxy1}, we obtain, first, the values for $x$ and $y$, and second, from \eqref{tauEetaubEe}, we obtain the stereometric coordinates
$\tau^E_e$ and $\taub^E_e$ such that
\begin{subequations}
\begin{align}
&\tau^E_e=\frac{P(\lambda,\taur'_5,\tan\alpha_e,\tan\beta_e)}{P_0(\lambda,\taur'_5,\tan\alpha_e,\tan\beta_e)}\,,\\
&\taub^E_e=\frac{\overline{P}(\lambda,\taur'_5,\tan\alpha_e,\tan\beta_e)}{P_0(\lambda,\taur'_5,\tan\alpha_e,\tan\beta_e)}\,,
\end{align}
\label{tauPQtabPbQb}
\end{subequations}
where  $P$, $\overline{P}$ and $P_0$  are polynomials of degree one with respect to $\lambda$ and $\taur'_5$ of which the coefficients are polynomials of degree one with respect to $\tan\alpha_e$ and $\tan\beta_e$.
\par
We also compute the four other pairs of time stamps ascribed to the event $e$, \ie, $(\taub^\Evb_e,\taut^\Evb_e)$, $(\taut^\Evt_e,\tauh^\Evt_e)$ and $(\tauh^\Evh_e,\tau^\Evh_e)$  (see Table \ref{tableattrib}), respectively, obtained at the events $\Evb$, $\Evt$ and $\Evh$. We obtain expressions similar to \eqref{tauPQtabPbQb} with respect to the other $\lambda$'s,  $\tau_5$'s,  $\tan\alpha$'s and  $\tan\beta$'s\,. And then, we set the following constraints:
\begin{equation}
	\begin{cases}
	\tau^E_e&=\tau^\Evh_e\,,\\
	\taub^E_e&=\taub^\Evb_e\,,\\
	\taut^\Evb_e&=\taut^\Evt_e\,,\\
	\tauh^\Evt_e&=\tauh^\Evh_e\,.
	\end{cases}
\label{secondtauEe}
\end{equation}
These constraints are well-justified because any event $e$ has only one 4-position. Then, we deduce four equations of the form
\begin{equation}
\lambda_1=\left(\frac{u\,\lambda_2+w}{w\,\lambda_2+r}\right),
\label{lambda1lambda2}
\end{equation}
for any pair $(\lambda_1,\lambda_2)$ of distinct $\lambda$ in the set $\{\lambda,\bar\lambda,\tilde\lambda,\hat\lambda\}$ from which we deduce one quadratic equation for each $\lambda$ with coefficients independent of the other
 $\lambda$'s but, nevertheless, depending on the angles and the various time stamps $\tau$. Therefore, we have proved that each $\lambda$ has a value which is independent on the other $\lambda$'s.
But, in addition, the $\lambda$'s must also be independent of the angles because  
they are ascribed to the projective points $[1,1]$ independently of the events such as $e$. Hence, we can  arbitrarily fix the values for the $\lambda$'s. The natural choice is to set the following:
\begin{equation}
\lambda\equiv\taur'_5\,,\qquad
\bar\lambda\equiv\taur^\bullet_5\,,\qquad
\tilde\lambda\equiv\taur^*_5\,,\qquad
\hat\lambda\equiv\taur^\circ_5\,.
\label{egaltau5lambdas}
\end{equation}
In return, from \eqref{lambda1lambda2} with \eqref{egaltau5lambdas}, we deduce also four fractional relations between, on the one hand, the $\alpha$'s,  and,  on the other hand, the $\beta$'s. The general form of these relations is the following. For instance, for $\tan\beta_e$, we have:
\begin{equation}
\tan\beta_e=\left(
\frac{u\,\tan\alpha_e+\bar{u}\,\tan\bar\alpha_e+\tilde{u}\,\tan\tilde\alpha_e+\hat{u}\,\tan\hat\alpha_e+r}{
w\,\tan\alpha_e+\bar{w}\,\tan\bar\alpha_e+\tilde{w}\,\tan\tilde\alpha_e+\hat{w}\,\tan\hat\alpha_e+s}
\right),
\end{equation}
where the coefficients $u$, $\bar u$, etc., depend on the time stamps except those ascribed to the localized event $e$.
\par
We then obtain the 4-position $p_e\equiv(\tau_e,\taub_e,\taut_e,\tauh_e)$ for $e$ in the grid such that $\tau_e\equiv\tau^E_e$, $\taub_e\equiv\taub^E_e$, $\taut_e\equiv\taut^\Evt_e$ and $\tauh_e\equiv\tauh^\Evt_e$ depending on the four angles $\alpha_e$, $\bar\alpha_e$, $\tilde\alpha_e$ and $\hat\alpha_e$ and the time stamps.
For instance, the stereometric coordinate $\tau_e$ satisfies
\begin{equation}
\tau_e=\left(
\frac{p\,\tan\alpha_e+\bar{p}\,\tan\bar\alpha_e+\tilde{p}\,\tan\tilde\alpha_e+\hat{p}\,\tan\hat\alpha_e+q}{
m\,\tan\alpha_e+\bar{m}\,\tan\bar\alpha_e+\tilde{m}\,\tan\tilde\alpha_e+\hat{m}\,\tan\hat\alpha_e+n}
\right).
\end{equation}
As a result, from 1) the form of this expression which is the same for each stereometric coordinate of the 4-position of $e$, and 2) following the same reasoning as in the preceding section for a $(2+1)$-dimensional spacetime, the group $PGL(5,\Rset)$ acts on $\Mmc$ via a projective transformation applied to the four tangents $\tan\alpha_e$, $\tan\bar\alpha_e$, $\tan\tilde\alpha_e$ and $\tan\hat\alpha_e$.
\par
Now, we can almost completely paraphrase what we described from p.~\pageref{tauialphajqell} in the preceding section, adding just one time stamp $\taub$ and another supplementary angle $\bar\alpha$. And then, following the same reasoning, we deduce that $\Mmc$ is modeled on $\Rset P^4$ and that it is embedded in $\Rset^5$. Finally, we denote by $\taur$ the fifth stereometric coordinate of the fibers of the submersion 
$\Rset^5$ to $\Mmc$. This supplementary stereometric coordinate $\taur$ is, actually, defined from the anchoring worldline $\mathring{W}$ following similarly the method indicated at the end of the last section.
\par\medskip
Lastly, the present protocol is based on the particular class of pairs of time stamps specified in the last right column of Table~\ref{tableattrib}. The stereometric coordinates ascribed to each event $e$ would differ for a different class of pairs.
Hence, we can obtain different, possible localizations for the same event $e$: a result which can be baffling only if we assume that localization is an absolute, intrinsic property of each spacetime event independent of any process.
But, after all, we are already faced with this situation when producing atlases of charts for manifolds. In the same way, we just need to know the changes of localization charts (stereometric grids) which are, actually, deduced naturally from the changes of charts defined by the changes of emission grids. Therefore, localization and location as well cannot be intrinsic processes.

\section{Conclusion \label{conclusion}}

Even though spacetime is represented by a four-dimensional manifold, the  localization processes show that spacetime cannot be physically apprehended if its manifold counterpart is not embedded geometrically in a five-dimensional manifold modeled locally on a four-dimensional projective space. Then, the spacetime manifold must be considered as a generalized Cartan manifold endowed necessarily, as a consequence, with a projective Cartan connexion.
\par
Also, the space and time splitting ascribed usually to the four dimensions of spacetime should be enhanced to encompass a fifth dimension. Then, to be complete, a supplementary notion should be added to space and time. We suggest the notion of energy.


\end{document}